\documentclass[aps,prb, amsmath,superscriptaddress,longbibliography,notitlepage]{revtex4-2}
\usepackage[latin9]{inputenc}
\usepackage{verbatim}
\usepackage{amsmath}
\usepackage{amssymb}
\usepackage{graphicx}
\usepackage{xcolor}
\usepackage{babel}

\begin{document}
\title{Semiclassical spin transport in LaO/STO system in the presence of multiple Rashba spin orbit couplings}
\author{Anirban Kundu}
\email{anirbank5@gmail.com}
\affiliation{Department of Electrical and Computer Engineering, National University of Singapore, Singapore 117583, Republic of Singapore}
\affiliation{Department of Physics, Ariel University, Ariel 40700, Israel}
\author{Zhuo Bin Siu}
\email{elesiuz@nus.edu.sg}
\affiliation{Department of Electrical and Computer Engineering, National University of Singapore, Singapore 117583, Republic of Singapore}
\author{Mansoor B.A. Jalil}
\email{elembaj@nus.edu.sg}
\affiliation{Department of Electrical and Computer Engineering, National University of Singapore, Singapore 117583, Republic of Singapore}

\begin{abstract}
The interaction between the linear and cubic spin-orbit coupling with magnetic moments and mobile spin-polarized carriers in the LaO/STO system provides new avenues for spin transport applications. We study the interplay between linear and cubic Rashba spin orbit coupling (RSOC) on in-plane magnetic moments in the LaO/STO system using the Boltzmann transport theory based on the relaxation time approximation (RTA) and the more refined Schliemann-Loss (SL) delta-potential scattering model. In general, both methods yield a linear (quadratic) relationship between the spin accumulation (spin current) when one of the three RSOC strengths is varied and the other two fixed. The simultaneous presence of multiple types of RSOC with distinct angular dependences is a key ingredient in breaking the $k$-space symmetry of the Fermi surface, thus ensuring a finite spin accumulation upon integration over the entire Fermi surface. While the oft-used RTA method is sufficiently accurate for spin accumulation calculations, the more refined SL model is required for spin current calculations because the RTA method neglects the anisotropy of the Fermi contour arising from the cubic RSOC terms. Based on the refined SL model and under optimal tuning of the RSOC parameters, the spin charge conversion values in LaO/STO is predicted to reach a remarkable efficiency of 30\%.
\end{abstract}
\maketitle
\section{Introduction}
A key requirement in spintronics devices is the efficient conversion of charge currents into spin accumulations or spin currents, which can be used to generate spin-transfer-torque to induce the magnetization switching in magnetic
memory systems based on magnetic multilayers, magnetic tunnel junctions
or magnetic skyrmions \cite{Nature425_6956,JMMM320_1190, NatMat5_210, Science330_1648, JAP110_121301, AnnPhys326_207, SGTbook}. Previous studies have demonstrated the efficacy of using Rashba spin-orbit coupling (RSOC) to achieve high charge-spin conversion efficiency in systems
with broken inversion symmetry \cite{Snchez2013,PhysRevLett.116.096602,Lesne2016,PhysRevLett.114.136402,PhysRevLett.117.076601,PhysRevB.97.041115,Ghiasi2019, PhysRevLett.112.096601,Yoda2018,PhysRevB.96.235419,Ghiasi2019,PhysRevB.93.195440,PhysRevApplied.13.054014,Salemi2019}.

Recently, oxide heterostructures have attracted much attention due
to their potential applications in spintronics devices \cite{doi:10.1063/1.4894640,Jin2017,Song2020,refId0,Songe1602312,Ohshima2017}.
Experimentally, oxide heterostructures have been shown to exhibit
superconducting \cite{Reyren1196}, and more importantly for our purpose,
ferromagnetic phases \cite{Ariando2011,Kalisky2012,PhysRevLett.107.096802}.
The surface of SrTiO$_{3}$ (STO) or its interface with other oxides
is also known to host two-dimensional electron gases (2DEGs) with high electron mobility \cite{Chen2013}.
The individual layers
in the oxide heterostructures are insulating, but when stacked together, they provide conduction electrons originating from the $d$-orbitals of the lattice system \cite{PhysRevLett.104.126803,Joshua2012,PhysRevB.87.161102}.
The carrier density and
related electronic properties can be strongly tuned using a gate voltage. This external is another key factor that renders the LaO/STO heterostructures as an attractive platform for
future spintronics architectures.

Our present work is motivated by the high spin-to-charge
 conversion efficiency in LaAlO$_3$/SrTiO$_3$ (LaO/STO) interfaces
under the influence of a strong SOC \cite{Wang2017,Trier2020,EPL116_17006}. The SOC experienced by conduction electrons in the 2DEG at the LaO/STO interface can be effectively described as a Rashba SOC effect due to the
broken inversion symmetry. The RSOC
in transition metal oxides \cite{PhysRevLett.112.086802,PhysRevB.95.165401,PhysRevLett.104.126803,PhysRevB.90.165108}
and electron-doped semiconductors \cite{PhysRevLett.78.1335,PhysRevB.62.4245,PhysRevLett.90.076807} has been shown to be linear in $\mathbf{k}$, i.e., having the generic form of $\alpha (\mathbf{k}\times e_{z})\cdot\boldsymbol{\sigma}$ where $e_{z}$ is the unit vector in the out-of-plane direction. However, in hole-doped semiconductors
\cite{PhysRevB.71.165312,PhysRevLett.113.086601,PhysRevLett.121.087701},
rare earth materials \cite{PhysRevLett.124.237202}, and in particular, at the surface
of STO \cite{PhysRevB.93.045108,PhysRevB.92.075309,PhysRevLett.108.206601}, the cubic RSOC also has a significant presence. In fact, the enhancement of the spin Hall conductivity in heavy-hole
semiconductors has been primarily attributed to the cubic RSOC \cite{PhysRevB.71.085308}.
In the LaO/STO system, the linear RSOC strength is approximately 10 meV\AA and that of cubic RSOC approximately 1--5 eV\AA$^3$ \cite{PhysRevLett.104.126803,PhysRevB.86.201105,PhysRevB.92.075309,Ho_2019}. Over the Brillouin zone of the system, the magnitudes of the linear and cubic SOC splittings in the LaO/STO system are comparable.

Owing to the key role of the RSOC in the charge transport of the LaO/STO system, it is important to go beyond the effective low-energy Hamiltonian to study the RSOC effect, and its dependence on the finite thickness and confining potential.  Such an investigation has been lacking even though there has been much study conducted on the spin accumulation and transport in the LaO/STO system \cite{APE15_013005, PhysRevResearch.3.013275, PhysRevB.102.144407, PhysRevLett.119.256801, PhysRevB.91.241302}. A previous study on the finite thickness of LaO/STO 2DEG system focuses on the out-of-plane transmission instead of transport within the 2D plane \cite{PhysRevResearch.3.043170}. To address the shortcomings in earlier derivations of effective low-energy Hamiltonians \cite{PhysRevB.86.125121, PhysRevB.87.161102} for the LaO/STO 2DEG, Ho et al. explicitly accounted for the spatial variation of the confining potential and the resultant quantum confinement \cite{Ho_2019} . In addition to the usual linear Rashba term in the effective Hamiltonian in the lowest-energy pair of spin-split bands, they found two additional cubic RSOC terms with the forms of $\beta_3(k_x^2-k_y^2)(k_y\sigma_x-k_x\sigma_y)$ and $\eta_3k_xk_y(k_x\sigma_x-k_y\sigma_y)$ where $\beta_3$ and $\eta_3$ are the strengths of the two cubic RSOC terms. The strengths of these cubic RSOC terms can be tuned by varying the thickness of the STO layer and the magnitude of the confining field. Although the effects of the $\beta_3$ term on the spin response in LaO/STO has been studied previously \cite{PhysRevB.91.241302,JMMM497_156919}, a full study of all the cubic RSOC terms in light of the above study has not been performed. In particular,  the $\eta_3$ term has a magnitude that can be comparable or even larger than that of the $\beta_3$ is significant with a magnitude comparable to or even exceeding that if the $\beta_3$ term, but has been neglected hitherto.

In this work, we investigate the interplay between the various linear and cubic RSOC effects on the spin accumulation and spin current in the presence of presence of an applied electric field and magnetization coupling. In our transport calculations, we use two approaches to obtain the non-equilibrium spin accumulation and current, namely, the conventional relaxation time approximation (RTA) and a more refined approach by Schliemann and Loss (SL) \cite{PhysRevB.68.165311}, which assumes delta-function scattering potentials and takes into account the anisotropy of the Fermi surface of the spin-split bands. We show that the simpler RTA approach is sufficient for the spin accumulation calculation but the more accurate SL approach is required to evaluate the spin current. Our results provide a more complete understanding of how the three RSOC terms a
ect the spin response of the LaO/STO system. We study the interplay between the various RSOC terms and their influence on the spin transport properties of the system. We find that under optimal tuning of the RSOC parameters, the LaO/STO system can reach high spin-charge conversion efficiency of about 30\%.

\section{Methods}

\begin{figure}[htp]
\centering
\includegraphics[width=0.4\textwidth]{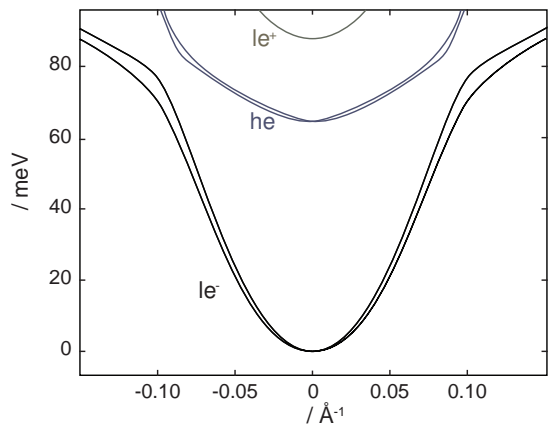}
\caption{  Band diagram of a typical LAO-STO quantum well system adopted from Fig. 1a of Ref. \cite{Ho_2019}. Here, we set the 0 energy level at the band bottom of the le$^-$ band.  }
\label{gFig1}
\end{figure}

The LaO/STO system possesses three pairs of conduction bands comprising the $d_{xy}$, $d_{yz}$, and $d_{zx}$ bands in which the inversion asymmetry across the interface results in a strong
spin-orbit interaction (SOC) in the system that results in energy splitting within each pair of bands. We adopt the Hamiltonian of Ho et al. \cite{Ho_2019}, who considered the lowest energy quantum well states for each of these bands due to the confining potential in the LaO/STO heterostructure. The resultant six bands in the 
in heterostructure can be then described by a six-by-six 
Hamiltonian representing two pairs of light electron bands denoted as le$^\pm$ and a pair of heavy electron bands $\mathrm{he}$ (Fig. 1a).   
Here, we restrict our consideration to low values of Fermi energies in which only the lowest-energy pair of spin-split bands, i.e., the $\mathrm{le}^-$ bands, are physically relevant. In the presence of magnetic
dpoants, the general form of the effective Hamiltonian for this pair of bands is, 
\begin{equation}
H=\frac{k^{2}}{2m^*}+J_{H}\boldsymbol{\sigma}\cdot\boldsymbol{M}+\alpha(\boldsymbol{\sigma}\times\boldsymbol{k})\cdot\hat{z}+\sigma_{x}\text{\ensuremath{\beta_{3}}}\left(k_{y}k_{x}^{2}-k_{y}^{3}\right)-\sigma_{x}\text{\ensuremath{\eta_{3}}}k_{x}^{2}k_{y}+\sigma_{y}\text{\ensuremath{\beta_{3}}}\left(k_{x}k_{y}^{2}-k_{x}^{3}\right)+\sigma_{y}\text{\ensuremath{\eta_{3}}}k_{y}^{2}k_{x}\label{eq:h-main}
\end{equation}
where $\alpha$ represents the strength of the linear Rashba SOC (RSOC), $\beta_{3}$
and $\eta_{3}$ are coefficients of the two distinct cubic SOC terms, and $J_{H}$
 the Heisenberg exchange interaction between conduction electrons
and the magnetic moments of the dopants $\boldsymbol{M}$. Here, we consider dopant magnetizations
lying in the in-plane direction with $\boldsymbol{M}= M_x\hat{x} + M_y\hat{y}$. Denoting the coefficients of $\sigma_x$ and $\sigma_y$ in Eq. \eqref{eq:h-main} as $h_x$ and $h_y$ respectively, $h_x$ and $h_y$ are explicitly given by 
\begin{align}
	h_x(\boldsymbol{k}) &= J_{H}M_{x}+\beta_{3}k_{y}\left(k_{x}^{2}-k_{y}^{2}\right) - \eta_{3}k_{x}^{2}k_{y}+\alpha k_{y} \label{hx} \\ 
	h_y(\boldsymbol{k}) &= J_{H}M_{y}-\beta_{3}k_{x}\left(k_{x}^{2}-k_{y}^{2}\right)+\eta_{3}k_{x}k_{y}^{2}-\alpha k_{x} \label{hy}.
\end{align} 
For later convenience, we also denote the spin-independent kinetic energy term as $h_0(\boldsymbol{k}) = \boldsymbol{k}^2/(2m^*)$. 
The eigenenergies of Eq. \eqref{eq:h-main} are then $\epsilon_{\boldsymbol{k}b}=h_0(\boldsymbol{k}) + b\sqrt{h_x^{2}\left(\mathrm{\boldsymbol{k}}\right)+h_y^{2}\left(\mathrm{\boldsymbol{k}}\right)}$ where $b=\pm 1$  are the indices for the spin polarized  bands and the corresponding eigenstates are given by
\begin{equation} 
\psi_{\boldsymbol{k}b}=\frac{1}{2\sqrt{A}}\left(\begin{array}{c}
\mathrm{e}^{-i\gamma_{\boldsymbol{k}}}\\
b
\end{array}\right)e^{i\boldsymbol{k}\cdot\boldsymbol{r}},
\end{equation}
where 
\begin{align}
\gamma_{\boldsymbol{k}} & =\tan^{-1}\left(\frac{h_y(\boldsymbol{k})}{h_x(\boldsymbol{k})}\right).\label{eq:gammak}
\end{align}
The spin $x$ expectation value of band $b$ is given by  
\begin{equation} 
	s^x_b = b\frac{h_x}{\sqrt{h_x^2+h_y^2}} 
\end{equation}
and the $x$-component of its group velocity given by 
\begin{equation}
v_{\boldsymbol{k}b}^{x}  = \frac{k_x}{m^*} + b \frac{ h_x\partial_{k_x}h_x(\boldsymbol{k}) + h_y\partial_{k_x}h_y(\boldsymbol{k})}{\sqrt{h_x(\boldsymbol{k})^2 + h_y(\boldsymbol{k})^2}}.\label{eq:vkx}
\end{equation}

With the above preliminary results for the spin expectation and group velocity values, we proceed to apply the Boltzmann transport equation (BTE) to calculate the non-equilibrium spin accumulation and spin current induced by an external electric field. The
general form of the BTE for each spin-split band is given by 
\begin{equation}
\frac{\partial  f_{\boldsymbol{k}b}(\boldsymbol{r},t)}{\partial t}+\dot{\boldsymbol{r}}\cdot\frac{\partial  f_{\boldsymbol{k}b}(\boldsymbol{r},t)}{\partial\boldsymbol{r}}+\dot{\boldsymbol{k}}\cdot\frac{\partial  f_{\boldsymbol{k}b}(\boldsymbol{r},t)}{\partial\boldsymbol{k}}=I_{\mathrm{coll}}\left\{ f_{\boldsymbol{k} b}(\boldsymbol{r},t)\right\} \label{eq:boltzmaneq1}
\end{equation}
where $f_{\boldsymbol{k}b}(\boldsymbol{r},t)$ is the non-equilibrium
distribution function of the $b$th band and $I_{\text{coll}}\left\{ f_{\boldsymbol{k}b}(\boldsymbol{r},t)\right\} $
is the collision integral. Without loss of generality, we choose the electric field $\boldsymbol{E}$ to be along the $x$ axis 
such
that $\dot{\boldsymbol{k}}=eE\hat{x}$. At steady
state, the explicit time derivative (the first term) is zero. Because we
consider a spatially homogenous system, we drop the
second term. Under the above two conditions, the BTE reduces to 
\begin{equation}
eE\frac{\partial f_{\boldsymbol{k}b}}{\partial k_{x}}=I_{\mathrm{coll}}\left\{ f_{\boldsymbol{k}b}\right\} .\label{eq:boltzmaneq2}
\end{equation}
Because Eq. \eqref{eq:boltzmaneq2} is a recursive equation, an exact general solution can only be obtained numerically. However, an analytical solution
can be obtained when the change in the distribution function is approximated
to the linear order in the external field. To solve Eq. \eqref{eq:boltzmaneq2}
analytically to the first order in the external field, we first linearize the distribution function as,

\begin{equation}
f_{\boldsymbol{k}b}=f_{\epsilon_{\boldsymbol{k}b}}^{(0)}+f_{\boldsymbol{k}b}^{(1)} \label{eq:fkb1},
\end{equation}
where $f_{\epsilon_{\boldsymbol{k}b}}^{(0)}$ is the equilibrium distribution, which 
depends only on the electron energy $\epsilon_{\mathrm{\boldsymbol{k}b}}$,
and the second term is the correction due to the external field. We
solve for $f_{\boldsymbol{k}b}$ via two approaches: (i)  the relaxation-time approximation (RTA) and (ii) a more refined method that considers the effect
of impurity scattering and the highly anisotropic nature of the dispersion using the method proposed
by Schliemann and Loss \cite{PhysRevB.68.165311}. Both approaches
will be discussed in detail in the next section. After solving the
BTE, we study the effect of the SOC on the transport coefficients comprising the spin accumulation, spin current, and charge-spin conversion efficiency.

The spin accumulation $\boldsymbol{S}$ is the net spin originating from the two spin-split
bands, 
\begin{align}
\boldsymbol{S} & =\sum_{b}\int\frac{\mathrm{d}\boldsymbol{k}}{\left(2\pi\right)^{2}}f_{\boldsymbol{k}b}^{(1)}\langle\psi_{\boldsymbol{k}b}|\boldsymbol{\sigma}|\psi_{\boldsymbol{k}b}\rangle\\
 & =\int\frac{\mathrm{d}\boldsymbol{k}}{\left(2\pi\right)^{2}}\left(f_{\boldsymbol{k}+}^{(1)}-f_{\boldsymbol{k}-}^{(1)}\right)\left(\hat{x}\cos\gamma_{\boldsymbol{k}}+\hat{y}\sin\gamma_{\boldsymbol{k}}\right)\label{eq:spin-accumulation}
\end{align}

The spin current density is the tensor product of both the velocity and spin of the electrons. A commonly adopted definition of the spin current is  \cite{PhysRevLett.96.076604}
\begin{equation}
	j_{\boldsymbol{v}, \boldsymbol{\sigma}}=\text{Re}\left( \psi^{\dagger}\frac{1}{2} \{ \boldsymbol{v},\boldsymbol{\sigma} \}\psi\right) \label{eq:OJs} 
\end{equation}

This gives, for $\psi = \psi_{\boldsymbol{k}b}$ and the spin current polarized in the spin $s=(x,y)$ direction flowing along the $l$ direction, 
\begin{equation}
	j_{ls} = b(\partial_{k_l}h_0)\frac{h_s}{\sqrt{h_x^2+h_y^2}} + \partial_{k_l}h_s \label{eq:Jis}.
\end{equation}

which then gives
\begin{align}
\delta J_{xx} & =\sum_{s}\int\frac{d\boldsymbol{k}}{\left(2\pi\right)^{2}}\left[ \left(f_{\boldsymbol{k}+}^{(1)}-f_{\boldsymbol{k}-}^{(1)}\right)(\partial_{k_x}h_0)\cos\gamma_{\boldsymbol{k}} + (f_{\boldsymbol{k}+}^{(1)} + f_{\boldsymbol{k}-}^{(1)})\partial_{k_x}h_x\right],\label{eq:spin-current-1}\\
\delta J_{xy} & =\sum_{s}\int\frac{d\boldsymbol{k}}{\left(2\pi\right)^{2}}\left[ \left(f_{\boldsymbol{k}+}^{(1)}-f_{\boldsymbol{k}-}^{(1)}\right)(\partial_{k_x}h_0)\sin\gamma_{\boldsymbol{k}}+ (f_{\boldsymbol{k}+}^{(1)} + f_{\boldsymbol{k}-}^{(1)})\partial_{k_x}h_y\right].\label{eq:spin-current-2}
\end{align}
(Here, the lower-case $j_{ls}$ denotes the expectation value of the spin current for a single state, and the uppercase $J_{ls}$ the non-equilibrium spin current obtained after integrating over all states. )

\subsection{Relaxation-time approximation (RTA)}

It is assumed in RTA that $I_{\mathrm{coll}}\left\{ f_{\boldsymbol{k}b}\right\} =f_{\boldsymbol{k}b}^{(1)}/\tau$
where $\tau$ is the relaxation time. Thus, from Eq. (\ref{eq:boltzmaneq2})
we obtain, 
\begin{equation}
f_{\boldsymbol{k}b}^{(1)}=-e\tau\left(\boldsymbol{E}\cdot\boldsymbol{v}_{\boldsymbol{k}b}\right)\left(-\frac{\partial f_{\epsilon_{\boldsymbol{k}b}}^{(0)}}{\partial\epsilon_{\boldsymbol{k}b}}\right).\label{eq:delf-constant-RT}
\end{equation}

At equilibrium, we assume that $f_{\epsilon_{\boldsymbol{k}b}}^{(0)}$ is given by the zero-temperature Fermi distribution function. As a consequence, $\left(-\partial f_{\epsilon_{\boldsymbol{k}b}}^{(0)}/\partial\epsilon_{\boldsymbol{k}b}\right)=\delta\left(\epsilon_{\boldsymbol{k}b}-\epsilon_{\mathrm{F}}\right)$
where $\epsilon_{\mathrm{F}}$ is the Fermi energy. Thus, only the electrons
at the Fermi level contribute to the transport. Accordingly, the change in the distribution function is obtained as
\begin{equation}
f_{\boldsymbol{k}b}^{(1)}  =-eE\tau v^x_{\boldsymbol{k}b}\delta(\epsilon_{\boldsymbol{k}b}-\epsilon_{\mathrm{F}}). \label{eq:f1RTA}
\end{equation}
With the above distribution function and Eq. (\ref{eq:spin-accumulation}),
the induced spin density due to the external electric field is given by the following integrals,

\begin{align}
S_{x}  =-eE\tau \int d\boldsymbol{k} &\left( \ \frac{h_x^2(\partial_{k_x}h_x) + h_xh_y(\partial_{k_x}h_y)}{h_x^2+h_y^2}(\delta(\epsilon_{\boldsymbol{k}+}-\epsilon_{\mathrm{F}})+\delta(\epsilon_{\boldsymbol{k}-}-\epsilon_{\mathrm{F}}))\right. \nonumber \\
&+ \left.\frac{ (\partial_{k_x}h_0)h_x}{\sqrt{h_x^2+h_y^2}}(\delta(\epsilon_{\boldsymbol{k}+}-\epsilon_{\mathrm{F}})-\delta(\epsilon_{\boldsymbol{k}-}-\epsilon_{\mathrm{F}}))\right)\\
S_{y}  =-eE\tau \int d\boldsymbol{k}( &\left( \ \frac{h_xh_y(\partial_{k_x}h_x) + h_xh_y^2(\partial_{k_x}h_y)}{h_x^2+h_y^2}\delta(\epsilon_{\boldsymbol{k}+}-\epsilon_{\mathrm{F}})+\delta(\epsilon_{\boldsymbol{k}-}-\epsilon_{\mathrm{F}}))\right. \nonumber \\
&+ \left. \frac{ (\partial_{k_x}h_0)h_y}{\sqrt{h_x^2+h_y^2}}(\delta(\epsilon_{\boldsymbol{k}+}-\epsilon_{\mathrm{F}})-\delta(\epsilon_{\boldsymbol{k}-}-\epsilon_{\mathrm{F}}))\right).
\end{align}
Note that in the above equations, because $\delta(f(x)) = \sum_{x_0} f(x_0)/|f'(x_0)|$ for an arbitrary function $f(x)$ where $x_0$ are the roots of  $f(x) $, 
\begin{equation}
	\int \mathrm{d}\boldsymbol{k}\ \delta(\epsilon_{\boldsymbol{k}b}-\epsilon_{\mathrm{F}})= \int_{\mathrm{FC}_b} \mathrm{d}l\ \frac{1}{v_{\boldsymbol{k}_{\mathrm{F}}b}} \label{eq:dkdelE}
\end{equation}
where $\int_{\mathrm{FC}_b} \mathrm{d}l$ indicates a line integral over the Fermi contour (FC) of band $b$, and the $v_{\boldsymbol{k}_{\mathrm{F}}b}^{-1}$ term reflects the density of states in the vicinity of the FC. We parameterize the points lying on the FC of band $b$ as $\boldsymbol{k}_{\mathrm{F}b} = k_{\mathrm{F}b}(\cos(\phi) \hat{x} + \sin(\phi)\hat{y})$ where $k_{\mathrm{F}b}$ at a particular $\phi$ is determined from the dispersion relation as the positive real solution to
\begin{align}
		\epsilon_{\mathrm{F}} = \frac{1}{2m^*}k_{\mathrm{F}b}^2 + b \sqrt{h_x(\boldsymbol{k}_{\mathrm{F}b})^2 + h_x(\boldsymbol{k}_{\mathrm{F}b})^2}
\label{eq:FC}.
\end{align}
Eq. \eqref{eq:dkdelE} gives
\begin{equation}
	\int \mathrm{d}\boldsymbol{k}\ \delta(\epsilon_{\boldsymbol{k}b}-\epsilon_{\mathrm{F}})= \int \mathrm{d}\phi \frac{k_{\mathrm{F}b}}{v_{\boldsymbol{k}_{\mathrm{F}}b}}. \label{intdkdel} 
\end{equation}

Similarly, from Eqs. \eqref{eq:spin-current-1} and \eqref{eq:spin-current-2},
the spin currents are given by the following integrals, 
\begin{align}
\delta J_{xx}  = -2eE\tau \int d\boldsymbol{k} \Bigg[  &\left((\partial_{k_x}h_0)(\partial_{k_x}h_x) + \frac{ (\partial_{k_x}h_0)h_x\left( (\partial_{k_x}h_x)h_x + (\partial_{k_x}h_y)h_y\right)}{h_x^2+h_y^2}\right)(\delta(\epsilon_{\boldsymbol{k}+}-\epsilon_{\mathrm{F}})+\delta(\epsilon_{\boldsymbol{k}-}-\epsilon_{\mathrm{F}})) \nonumber \\
&\left( \frac{h_x( (\partial_{k_x}h_0)^2 + (\partial_{k_x}h_x)^2) + (\partial_{k_x}h_x)(\partial_{k_x}h_y)h_y}{\sqrt{h_x^2+h_y^2}} \right)(\delta(\epsilon_{\boldsymbol{k}+}-\epsilon_{\mathrm{F}})-\delta(\epsilon_{\boldsymbol{k}-}-\epsilon_{\mathrm{F}})) \Bigg]\\
J_{xy}  = -2eE\tau \int d\boldsymbol{k} \Bigg[  &\left((\partial_{k_x}h_0)(\partial_{k_x}h_y) + \frac{ (\partial_{k_x}h_0)h_y\left( (\partial_{k_x}h_x)h_x + (\partial_{k_x}h_y)h_y\right)}{h_x^2+h_y^2}\right)(\delta(\epsilon_{\boldsymbol{k}+}-\epsilon_{\mathrm{F}})+\delta(\epsilon_{\boldsymbol{k}-}-\epsilon_{\mathrm{F}})) \nonumber \\
&\left( \frac{h_y( (\partial_{k_x}h_0)^2 + (\partial_{k_x}h_x)^2) + (\partial_{k_x}h_x)(\partial_{k_x}h_y)h_x}{\sqrt{h_x^2+h_y^2}} \right)(\delta(\epsilon_{\boldsymbol{k}+}-\epsilon_{\mathrm{F}})-\delta(\epsilon_{\boldsymbol{k}-}-\epsilon_{\mathrm{F}})) \Bigg].
\end{align}
The integrations over the FCs are evaluated numerically to obtain the 
values of the spin densities and currents at various SOC strengths.

\subsection{Schliemann-Loss approach}

In the more refined approach of Schliemann and Loss \cite{PhysRevB.68.165311}, we assume that scattering occurs due to impurities modeled as delta potential functions. In this method, which we shall henceforth refer to as the Schliemann-Loss (SL) approach  for short, the collision
integral is given by \cite{0030839939} 
\begin{equation}
I_{\textrm{coll}}\left\{ f_{\boldsymbol{k}b}\right\} =\int \mathrm{d}\boldsymbol{k}' \sum_{b,b'}\mathrm{w}_{\boldsymbol{k}b,\boldsymbol{k}'b'}\left(f_{\boldsymbol{k}b}-f_{\boldsymbol{k}'b'}\right) \label{eq:SLIcoll1}
\end{equation}
where $\mathrm{w}_{\boldsymbol{k}s,\boldsymbol{k}'s'}$ is the transition probability for electrons in the $(\boldsymbol{k},s)$ state to be scattered into the $(\boldsymbol{k}', s')$ state  and only the linear terms in the applied electric field are retained in $f_{\boldsymbol{k}b}$ as per Eq. \eqref{eq:fkb1}.  
We consider electrons to be scattered by impurities. Each impurity is modeled as a delta
function potential given by 
\begin{equation}
	V_i(\boldsymbol{r})=a_0V_{0}\delta(\boldsymbol{r}-\boldsymbol{R}_{i}) \label{eq:Vr}
\end{equation}
where $\boldsymbol{R}_{i}$ is the position of the impurity and $V_{0}$ its strength. 
In the above, $a_0$ is a quantity with the physical dimensions of area introduced for dimensional consistency noting that $\delta(\boldsymbol{r})$ has the dimensions of inverse area because  $\int \mathrm{d}\boldsymbol{r} \delta(\boldsymbol{r}) = 1$. The meaning of $a_0$ can be understood intuitively as follows : For an arbitrary potential $V'(\boldsymbol{r})$ with a finite spatial extent rather than a delta potential, $\int \mathrm{d}\boldsymbol{r}\ V'(\boldsymbol{r})$ gives a quantification of the influence of the potential over the entire system, Hence, $a_0V_0$ in Eq. \eqref{eq:Vr}, which is also the result of integrating Eq. \eqref{eq:Vr} over the entire system, is the corresponding influence of each of the delta-potential scatterers.  

Denoting the number density (i.e., the number per unit area) of such impurities as $n$ and applying the Fermi golden rule, the transition rate due to all the impurities is given by  
\begin{align}
\mathrm{w}_{\boldsymbol{k}b,\boldsymbol{k}'b'} & = n \frac{2\pi}{\hbar}\left|\langle\boldsymbol{k}b|a_0V_0|\boldsymbol{k}'b'\rangle\right|^{2}\delta\left(\epsilon_{\boldsymbol{k}b}-\epsilon_{\boldsymbol{k}'b'}\right) \label{eq:wkkprime1} \\
 & =\frac{1}{2} w_0 \left(1+bb'\cos(\gamma_{\boldsymbol{k}}-\gamma_{\boldsymbol{k}'})\right)\delta\left(\epsilon_{\boldsymbol{k}s}-\epsilon_{\boldsymbol{k}'b'}\right),\label{eq:wkkprime2}
\end{align}
(here, we have temporarily restored the $\hbar$ as a check of the dimensional consistency of $w$, and introduced $w_0 \equiv 2\pi n(a_0V_0)^2/\hbar$, which has the physical dimensions of [area][energy]/[time] so that $\boldsymbol{w}$ has the dimensions of [area]/[time]. This is required by the dimensional consistency of Eq. \eqref{eq:boltzmaneq2}, where $I_{\mathrm{coll}}$ is required to have dimensions of 1/[time], and that of Eq. \eqref{eq:SLIcoll1}, where $\boldsymbol{w}$ is consequently required to have the dimensions of [area]/[time]. Note that the factor of area appears to be misplaced in Eq. (38) in the original Schliemann and Loss paper, which corresponds to Eq. \eqref{eq:wkkprime1} here. ) 


The solution for $f_{\boldsymbol{k}b}^{(1)}$ in the BTE with the collision integral given in Eq. \eqref{eq:SLIcoll1} can be decomposed into
longitudinal and transverse components as follows, 
\begin{equation}
f_{\boldsymbol{k}b}^{(1)}=f_{\boldsymbol{k}b}^{(1)\parallel}+f_{\boldsymbol{k}b}^{(1)\perp}.\label{eq:delf}
\end{equation}
Both terms are independent of each other and defined as follows,
\begin{equation}
f_{\boldsymbol{k}b}^{(1)\parallel}=-e\left(\boldsymbol{E}\cdot\boldsymbol{v}_{\boldsymbol{k}b}\right)\left(-\frac{\partial f_{\epsilon_{\boldsymbol{k}b}}^{(0)}}{\partial\epsilon_{\boldsymbol{k}b}}\right)\left[\frac{\frac{1}{\tau_{\boldsymbol{k}b}^{\parallel}}}{\left(\frac{1}{\tau_{\boldsymbol{k}b}^{\parallel}}\right)^2+\left(\frac{1}{\tau_{\boldsymbol{k}b}^{\perp}}\right)^{2}}\right],\label{eq:delfparallel}
\end{equation}
\begin{align}
f_{\boldsymbol{k}b}^{(1)\perp} & =-e\left(\hat{\boldsymbol{z}}\times\boldsymbol{E}\cdot\boldsymbol{v}_{\boldsymbol{k}b}\right)\left(-\frac{\partial f_{\epsilon_{\boldsymbol{k}b}}^{(0)}}{\partial\epsilon_{\boldsymbol{k}b}}\right)\left[\frac{\frac{1}{\tau_{\boldsymbol{k}b}^{\perp}}}{\left(\frac{1}{\tau_{\boldsymbol{k}b}^{\perp}}\right)^2+\left(\frac{1}{\tau_{\boldsymbol{k}b}^{\parallel}}\right)^{2}}\right] \label{eq:delfperp}
\end{align}
where the (inverses of the) longitudinal relaxation time $\tau_{\boldsymbol{k},b}^{\parallel}$
and the transverse relaxation time $\tau_{\boldsymbol{k},b}^{\perp}$
are, 
\begin{align}
\frac{1}{\tau_{\boldsymbol{k}b}^{\parallel}} & =\sum_{b'}\int\frac{\mathrm{d}\boldsymbol{k}'}{\left(2\pi\right)^{2}}\mathrm{w}_{\boldsymbol{k}b,\boldsymbol{k}'b'}\left(1-\frac{|\boldsymbol{v}_{\boldsymbol{k}'b'}|}{|\boldsymbol{v}_{\boldsymbol{k}b}|}\cos\left(\vartheta_{\boldsymbol{k}b}-\vartheta_{\boldsymbol{k}'b'}\right)\right)\label{eq:relaxation-time-1}\\
\frac{1}{\tau_{\boldsymbol{k}b}^{\perp}} & =\sum_{b'}\int\frac{\mathrm{d}\boldsymbol{k}'}{\left(2\pi\right)^{2}}\mathrm{w}_{\boldsymbol{k}b,\boldsymbol{k}'b'}\frac{|\boldsymbol{v}_{\boldsymbol{k}'b'}|}{|\boldsymbol{v}_{\boldsymbol{k}b}|}\sin\left(\vartheta_{\boldsymbol{k}b}-\vartheta_{\boldsymbol{k}'b'}\right)\label{eq:relaxation-time-2}.
\end{align}
In the above equations, $\vartheta_{\boldsymbol{k}b}$ is the angle between $\boldsymbol{v}_{\boldsymbol{k}b}(=\partial\epsilon_{\boldsymbol{k}b}/\partial\boldsymbol{k})$
and $\boldsymbol{E}$, i.e., $\vartheta_{\boldsymbol{k}b}=\angle\left(\boldsymbol{v}_{\boldsymbol{k}b},\boldsymbol{E}\right)$.

Eqs. \eqref{eq:spin-accumulation}, \eqref{eq:spin-current-1}, and \eqref{eq:spin-current-2} are then evaluated numerically using $f_{\boldsymbol{k}s}^{(1)}$ in Eq. \eqref{eq:delf} to obtain the spin accumulations and currents.

\section{Results and discussion} 
For all our numerical results that follow, we set $\epsilon_{\mathrm{F}} = 30$ meV from the band bottom of the le$^-$ band and $J_HM=10$ meV. These values are chosen so that the Fermi energy falls around the middle of the energy range in which only the le$^-$ spin-split bands exist as propagating states in the bandstructure of Fig. \ref{gFig1} and the Fermi energy is sufficiently far from the band bottoms of the he and le$^+$ bands for the effects of these bands to be negligible. We treat the SOC strengths $\alpha$, $\beta_3$, and $\eta_3$ in Eq. \eqref{eq:h-main} as free parameters and vary $\alpha$ from -10 to 10 meV\AA$^{-1}$, and $\beta_3$ and $\eta_3$ from -3 to 3 eV\AA$^{-3}$. These ranges for the SOC strengths are based on the values in Ref. \cite{Ho_2019} corresponding to typical thicknesses and out-of-plane electric fields in LAO-STO quantum well structures. 

Furthermore, we note that because of the $\delta(\epsilon_{\mathbf{k},b}-\epsilon_{\mathrm{F}})$ term in  Eq. \eqref{eq:f1RTA} for the RTA approach and in Eq. \eqref{eq:wkkprime2} for the SL approach and Eq. \eqref{intdkdel}, the non-equilibrium quantity $O\in\{S,J_{ls}\}$ can be written in the generic form of
\begin{equation}
	O = E \sum_{b=\pm} \int \mathrm{d}\phi\ \Phi_b(\phi) o_b(\phi) \label{eq:PhiB}
\end{equation}
where $o_b(\phi)$ is $\boldsymbol{s}$ or $j_{ls}$ for a state at the polar angle $\phi$ on the Fermi contour of band $b$, and $\Phi_b(\phi)$ is a pre-factor that is proportional to the constant $\tau$ in the RTA approach and to the constant $1/(n(a_0V_0)^2)$ in the SL approach. This implies that the RTA and SL results for a given observable for a given set of parameters identical to each other when $\tau$ in the former and $1/(n(a_0V_0)^2)$ in the latter correspond to each other in the following sense: Since the magnitude of $(\tau^\parallel_{\boldsymbol{k}s})^{-1}$ in the SL approach is much larger than that of $(\tau^\perp_{\boldsymbol{k}s})^{-1}$ in practice  (see Supplementary Materials Fig. S1), the latter can be approximated to 0. In this case, the contribution of $f^{(1)\perp}_{\mathbf{k}b}$ in Eq. \eqref{eq:delfperp} becomes 0 while $f^{(1)\parallel}_{\mathbf{k}b}$ becomes proportional to $\tau^\parallel_{\boldsymbol{k}b}$ and has a similar form to the RTA $f^{(1)\parallel}_{\mathbf{k}b}$ in Eq. \eqref{eq:delf-constant-RT} except for its $\boldsymbol{k}$ and $b$ dependence of $\tau^\parallel_{\boldsymbol{k}b}$ in the former. The value of $\tau$ in the RTA approach can thus be considered as the average of $\tau^\parallel_{\boldsymbol{k}b}$ f the Fermi contours of the two spin-split bands  for a given value of $1/(n(a_0V_0)^2)$ in the SL approach.      

In the numerical results that follow, we assume that this correspondence between the RTA $\tau$ and SL $1/(n(a_0V_0)^2)$ for ease of comparison between the RTA and SL results. 

\subsection{SOC terms} 
\begin{figure}[htp]
\centering
\includegraphics[width=0.9\textwidth]{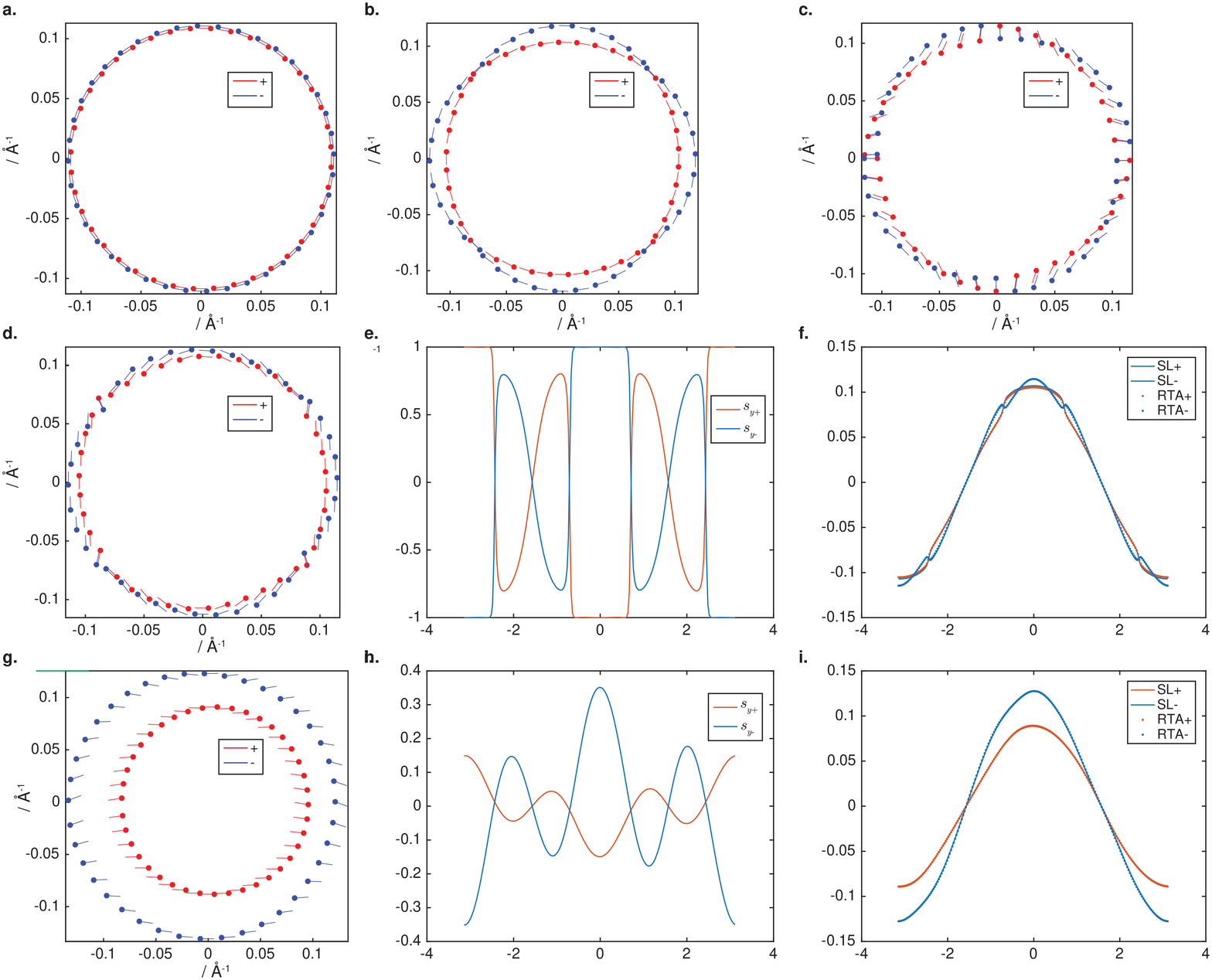}
\caption{\textbf{a}--\textbf{d} Fermi contours and spin expectation value vector of each state on the Fermi contours at $\boldsymbol{M}=0$ and a., $\alpha=5$ meV\AA$^{-1}$, $\beta_3=\eta_3=0$ eV\AA$^{-3}$ ; b. $\alpha=0$ eV\AA$^{-1}$, $\eta_3=0$ eV\AA$^{-3}$, $\beta_3=3$ eV\AA$^{-3}$;   c. $\alpha=0$ eV\AA$^{-1}$, $\beta_3=0$ eV\AA$^{-3}$, $\eta_3=3$ eV\AA$^{-3}$; and d.  $\alpha=5$ meV\AA$^{-1}$,and $\beta_3=\eta_3=1.5$ eV\AA$^{-3}$. The red and red arrows correspond to the $+$ and $-$ bands, respectively. \textbf{e} shows the spin $y$ expectation value $s_y$  for the parameter set of d. as functions of the wavevector angle $\phi$, and \textbf{f} the corresponding $\Phi_b(\phi)$ for the RTA and SL approaches for the two bands. \textbf{g}--\textbf{i} show the corresponding Fermi contours, $\Phi_b$, and spin $x$ expectation values for the same parameter set as d. except that $J_HM$ now has a finite value of $J_HM=10$ meV. }
\label{gFig2}
\end{figure}

We first introduce the effects of the various SOC terms in the Hamiltonian Eq. \eqref{eq:h-main}. Figure \ref{gFig2}a, b, and c show the respective Fermi contours of the two spin-split bands and the spin expectation value vectors of each state on the Fermi contours when $J_HM=0$ and \textcolor{blue}{the hypothetical scenarios where} only one of $\alpha$, $\beta_3$, or $\eta_3$ has a finite value. 

The system reduces to the familiar linear Rashba 2DEG when only $\alpha$ has a finite value while $J_HM$,  $\beta_3$, and $\eta_3$ are all zero. The contribution $H_\alpha$ of the linear RSOC term to the Hamiltonian Eq. \eqref{eq:h-main} in polar coordinates is,
\begin{equation}
	H_\alpha = \alpha(\sin(\phi)\sigma_x - \cos(\phi)\sigma_y). \label{eq:Ha}
\end{equation}
  As shown in Fig. \ref{gFig2}a, the corresponding Fermi contours comprise of two concentric circles with constant radii. The spin expectation value vectors are tangential to the Fermi contours and rotate in a clockwise (anti-clockwise) manner around the Fermi cotour for the + (-) band. The variation of $k_{\mathrm{F}b}$ and the spin expectation value vectors is more complicated for the cubic SOC terms. 

We next consider the terms proportional to $\beta_3$ in Eq. \ref{eq:h-main}. Fig. \ref{gFig2}b shows that when only $\beta_3$ is non-zero, the spin expectation vectors are approximately tangential to the Fermi contour reminiscent of the linear Rashba 2DEG. In contrast to the Rashba 2DEG, the radii of the Fermi contours now vary with $\phi$. In particular, the Fermi contours for the two bands cross each other at $\phi=\pm \pi/4, \pi\pm\pi/4$, as can be determined from the polar coordinates form of the contribution $H_{\beta_3}$ of the terms proportional to the Hamiltonian $\beta_3$to Eq. \eqref{eq:h-main},
\begin{equation}
H_{\beta_3} = \beta_3k^3\cos(2\phi) ( \sin(\phi)\sigma_x - \cos(\phi)\sigma_y). \label{eq:Hb3}
\end{equation}
The rotation of the spin expectation vectors in each of the bands switches directions at the values of $\phi$ where the two bands touch each other.  

Fig. \ref{gFig2}c shows the Fermi contours and the spin expectation vectors when only $\eta_3$ has a non-zero value. The contribution $H_{\eta_3}$ of the $\eta_3$ terms to the Hamiltonian Eq. \eqref{eq:h-main} in polar coordinates form is 
\begin{equation}
	H_{\eta_3} = \eta_3k^3(-\cos^2(\phi)\sin(\phi)\sigma_x+\sin^2(\phi)\cos(\phi)\sigma_y). \label{eq:He3}
\end{equation}
Eq. \ref{eq:He3} implies that the two bands cross each other at $\phi=0, \pm \pi/2, \pi$. In contrast to the linear RSOC and $\beta_3$ SOC for which the spin expectation value vectors are  tangential to the Fermi contours, the spin expectation vectors are normal to the Fermi contours at these band touching points. Similar to the $\beta_3$ SOC term, the rotation of the spin expectation value vectors in each band reverses direction at the values of $\phi$ where the two bands touch each other. 

Eqs. \eqref{eq:Ha}--\eqref{eq:He3} imply that the spin $x$ expectation values are antisymmetric about the $k_x$ axis and symmetric about the spin $k_y$ axis, whereas the spin $k_y$ expectation values are symmetric about the $k_x$ axis and antisymmetric about the $k_y$ axis when the linear Rashba and the cubic $\eta_3$ and $\beta_3$ SOCs are individually present. The same symmetries are therefore present when more than one of these SOC terms are simultaneously present in the system, as shown in Fig. \ref{gFig2}d for both in-plane spin directions and in Fig. \ref{gFig2}e for spin $y$ specifically when all three SOC terms have finite values. Fig. \ref{gFig2}d shows that the Fermi contours retain the reflection symmetries about the $k_x$ and $k_y$ axes although the band-touching points are now shifted away from $\phi=0$,$\pi$,$\pm\pi/4$,$\pi$, or $\pm \pi/4$. (The shift can also be seen from the values of $\phi$ at which $s_y=0$ in Fig. \ref{gFig2}e besides $\phi=\pm\pi/2$, at which the spin expectation vector lies along the $\pm x$ directions. ) The preserved antisymmetry of $s_y$ about $\phi=0$ is clearly evident in Fig. \ref{gFig2}e.

Fig. \ref{gFig2}f shows the factor $\Phi_b(\phi)$ in Eq. \eqref{eq:PhiB} obtained using the SL and RTA approaches for the parameter set of Fig. \ref{gFig2}e. When the RTA $\tau$ and SL $1/(n(a_0V_0)^2)$ are in correspondence such that the maximum values of $\Phi_b(\phi)$ for both approaches match, the $\Phi_b(\phi)$ profiles for the two approaches are similar but not exactly identical to each other. (The differences between the RTA and SL profiles cannot be clearly seen at the scale of Fig. \ref{gFig2}e. However, these differences do lead to visible differences in the variational trends of the spin currents. ) Because of the symmetries of the SOC terms and the resultant symmetries of the Fermi contours, $\Phi_b(\phi)$ is symmetrical about $\phi=0$. 

The aforementioned symmetries are broken by the application of the external magnetization, which breaks the time-reversal symmetry of the system. Fig. \ref{gFig2}g shows the Fermi contours and spin expectation value vectors for the same parameter set as Fig. \ref{gFig2}d except that a finite magnetization is now applied along the $x$ direction. Unlike the previous cases considered in which the Fermi contours for the two bands are both centered on the $k$-space origin, the centers of the Fermi contours are now slightly displaced with respect to each other. This is because the change in the spin degree of freedom due to the magnetization affects the momentum degree of freedom due to the SOC terms that couple the spin and momentum degrees of freedom together. The broken symmetries are also evident from the spin $y$ expectation values plotted in Fig. \ref{gFig2}h where the $s_y$ peaks at negative values of $\phi$ are slightly smaller than those at positive values of $s_y$. $\Phi_b(\phi)$ plotted in Fig. \ref{gFig2} also has an evident asymmetry about $\phi=0$. However, the $\Phi_b(\phi)$ profiles obtained using the SL and RTA approaches are still very similar to each other. Recall from Eq. \ref{eq:PhiB} that the non-equilibrium spin $y$ accumulation $\delta S_y$ is proportional to the product of $s_y(\phi)$ (Fig. \ref{gFig2}h) and $\Phi_b(\phi)$ summed over $b=\pm$ and integrated over $\phi$ from $-\pi$ to $\pi$. The asymmetry of $s_y(\phi)$ and $\Phi_b(\phi)$ with respect to both $\phi$ and $b$ reduces the cancellation between the contributions of the two bands to $\delta S_y$ and increases the amplitude of $\delta S_y$ compared to the case where the magnetization is absent. (The contributions of the two bands in Fig. \ref{gFig2}d--f without the magnetization cancel each other out to a relatively larger degree but not completely because of the differences in the $\Phi_b(\phi)$ profile between $b=\pm$ in Fig. \ref{gFig2}f.) It should also be noted that when only the $x$ magnetization is present but not SOC, $\delta S_y$ is zero because $s_{yb}(\phi)=0$ for all values of $\phi$ and $b$. The SOC results in finite values of $s_y$, which becomes asymmetrical (Fig. \ref{gFig2}h) when the $x$ magnetization is applied. This point will be relevant to the subsequent results where we study the effects of varying the three SOC strengths $\alpha$, $\beta_3$, and $\eta_3$ on the spin accumulation and spin current.  

\subsection{Spin accumulation} 
Noting that only the spin accumulation polarized perpendicular to the applied magnetization will exert a spin torque on the magnetization, we first study the spin accumulation component in the perpendicular direction to the magnetization. Writing the magnetization vector $\boldsymbol{M}=M\hat{m}$ with $\hat{m}=\cos(\phi_{M})\hat{x} + \sin(\phi_M)\hat{y}$ where $\phi_{M}$ is the magnetization angle, we denote the spin polarization in the $\hat{z}\times\hat{m}$ direction as $\delta S_\perp$ and study its variation with the SOC strengths.

\begin{figure}[htp]
\centering
\includegraphics[width=0.9\textwidth]{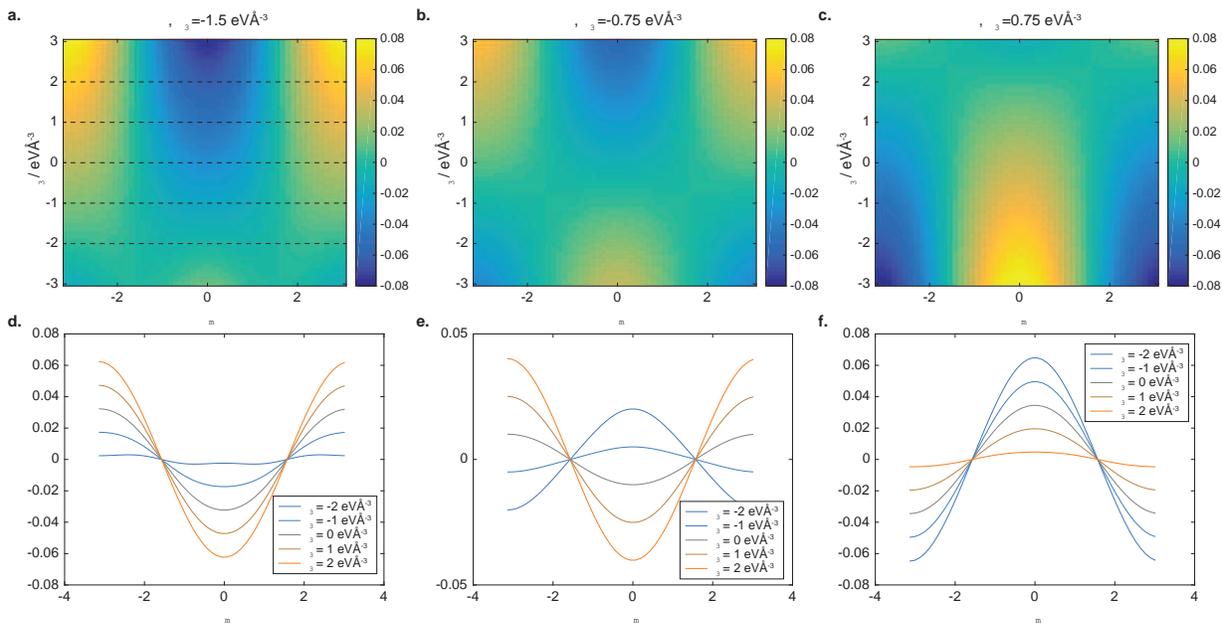}
\caption{\textbf{a}--\textbf{c} $\delta S_\perp$ as a function of the magnetization angle $\phi_{M}$ and $\eta_3$ for $\alpha=5$ meV\AA$^{-1}$ and a. $\beta_3=-1.5$ eV\AA$^{-3}$, b. $\beta_3=-0.75$ eV\AA$^{-3}$, and c. $\beta_3=0.75$ eV\AA$^{-3}$. \textbf{d}--\textbf{e} show the values of $\delta S_\perp$ plotted as functions of $\phi_{M}$ at various values of $\eta_3$ indicated by the dotted lines in a. for the parameter sets in a.-c., respectively. }

\label{gFig3}
\end{figure}	

Fig. \ref{gFig3} shows $\delta S_\perp$ plotted as functions of the magnetization angle $\phi_{M}$ and $\eta_3$ at a fixed value of $\alpha$ while varying the value of $\beta_3$. The variation of $\delta S_\perp$ with $\phi_m$ at the specific values of $\eta_3$ corresponding to the horizontal dotted lines in Fig. \ref{gFig3}a are plotted out in the graphs on the lower row of the figure to provide a better visualization of the variation of $\delta S_\perp$.  For a given set of $\alpha$, $\beta_3$, and $\eta_3$, $\delta S_\perp$ varies approximately as $\sin(\phi_m)$ . (This variation is not exact - for example, a significant deviation from a $\sin(\phi_m)$ behavior can be seen in the plot corresponding to $\eta_3=-2$ eV\AA$^{-3}$ line in Fig. \ref{gFig3}d in which there are three local maxima as $\phi_m$ varies from $-\pi$ to $\pi$.)  $\delta S_\perp$ has the largest magnitude at a given set of $\alpha$, $\beta_3$, and $\eta_3$ when the magnetization is parallel or anti-parallel to the applied electric field in the $x$ direction ($\phi_{M}=0,\pi$) and is identically 0 when the magnetization is perpendicular to the applied electric field ($\phi_{M}=\pm\pi/2$). As $\eta_3$ increases from a large negative value to a less negative value, the amplitude of $\delta S_\perp$ decreases until it reaches almost zero across a transitional range of $\eta_3$ (e.g., the $\eta_3=-2$ eV\AA$^{-3}$ line in Fig. \ref{gFig3}d). A further increase in $\eta_3$ causes the sign of $\delta S_\perp$ to flip (e.g., the $\eta_3=0$ eV\AA$^{-3}$ line vs the $\eta_3=1$ eV\AA$^{-3}$ line in Fig. \ref{gFig3}e). The amplitude of $\delta S_\perp$ then continues to increase in the reverse directionas $\eta_3$ increases further.  The value of $\eta_3$ at which $\delta S_\perp$ flips sign is dependent on the value of $\beta_3$  - it occurs at approximately $\eta_3=-2.5$ eV\AA$^{-3}$ in Fig. \ref{gFig3}a where $\beta_3=-1.5$ eV\AA$^{-3}$, $\eta_3=-0.7$ in Fig. \ref{gFig3}b where $\beta_3=-0.75$ eV\AA$^{-3}$, and $\eta_3=2.1$ in Fig. \ref{gFig3}c where $\beta3=0.75$ eV\AA$^{-3}$. A comparison between Figs. \ref{gFig3}a and b and Fig. \ref{gFig3}c shows that varying the value of $\beta_3$ can also flip the sign of $\delta S_\perp$.  Since the largest amplitude of $S_{\perp}$ is usually obtained when the magnetization is applied along the $\pm x$ directions, we now focus on $\delta S_y$ in the subsequent discussion. 

\begin{figure}[htp]
\centering
\includegraphics[width=0.7\textwidth]{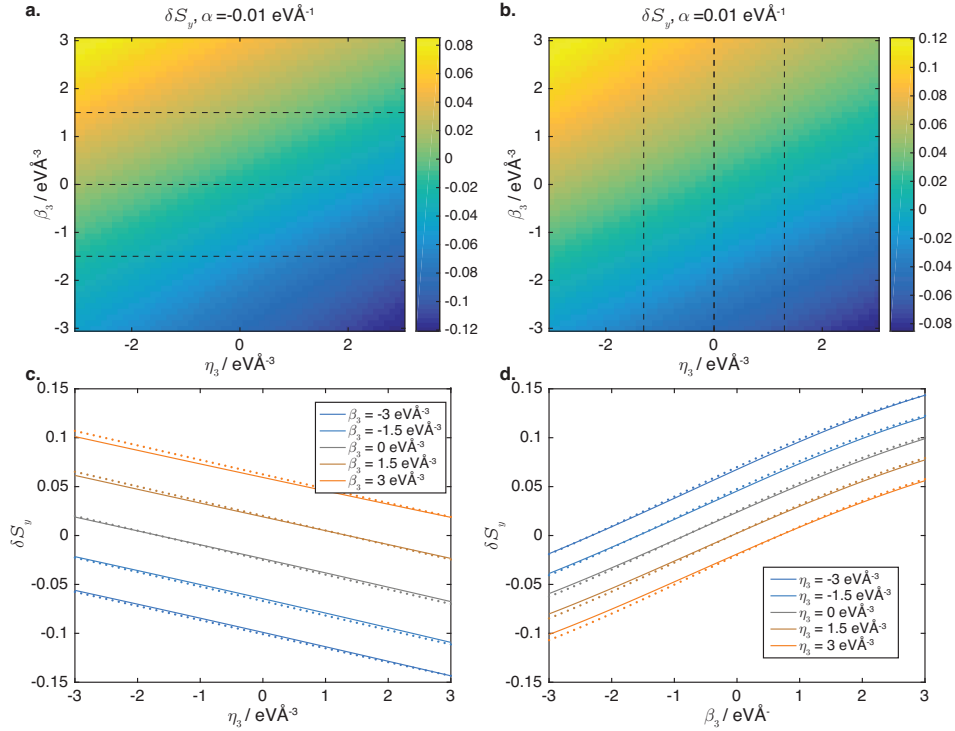}
\caption{
\textbf{a}--\textbf{b} Spin $y$ accumulation $\delta S_y$ as a function of $\eta_3$ and $\beta_3$ at a. $\alpha=-0.01$ eV\AA$^{-1}$ and b. $\alpha=0.01$ eV\AA$^{-1}$. \textbf{c} $\delta S_y$ as a function of $\eta_3$ for the parameters in a. at the various values of $\beta_3$ indicated by the dotted lines in a., and \textbf{d} $\delta S_y$ as a function of $\eta_3$ for the parameters in b. at the various values of $\eta_3$ indicated by the dotted lines in b. The solid lines in c. and d. denote $\delta S_y$ obtained using the SL approach, and the dots $\delta S_y$ obtained using RTA. }
\label{gFig4}
\end{figure}	

Fig. \ref{gFig4}a and b show $\delta S_y$ obtained using the SL approach plotted as functions of $\eta_3$ and $\beta_3$ for two different values of $\alpha$. The values of $\delta S_y$ at the specific values of $\beta_3$ indicated by the horizontal dotted lines in Fig. \ref{gFig4}a and at the specific values of $\eta_3$ indicated by the vertical dotted lines in Fig. \ref{gFig4}b are respectively plotted as functions of $\beta_3$ and $\eta_3$ in Fig. \ref{gFig4}c and \ref{gFig4}d to provide a clearer visualization of the variational trends along with the corresponding values of $\delta S_y$ obtained using the RTA approach. Fig. \ref{gFig4}a and \ref{gFig4}b show that $\delta S_y$ varies monotonically with $\eta_3$ and $\beta_3$. Fig. \ref{gFig4}c and \ref{gFig4}d respectively show more clearly that $\delta S_y$ has an approximately linear dependence on $\eta_3$ and $\beta_3$. Fig. S2 in the Supplementary Materials, which shows analogous plots of $\delta S_y$ as functions of $\alpha$ and $\eta_3$ at two values of $\beta_3$, indicates that $\delta S_y$ varies approximately linearly with $\alpha$ as well. This linear dependence of $\delta S_y$ on $\alpha$ is reflected in the fact although the color plots in Fig. \ref{gFig4}a and \ref{gFig4}b are visually similar, their color scale bars span different ranges. The approximately linear dependence of $\delta S_y$ on $\alpha$, $\beta_3$, and $\eta_3$ implies that there exist critical values of one of these three parameters at which $\delta S_y$ changes sign when the other two SOC parameters are fixed, as we have seen earlier in the discussion on Fig. \ref{gFig3} in which $\eta_3$ was varied while $\beta_3$ and $\alpha$ were fixed. 

These trends for the variation of one of the three SOC (say, $\eta_3$) while the other two SOC fields ($\alpha$,$\beta_3$) are fixed can be explained as follows: The spin-dependent parts of the Hamiltonian in Eq. \eqref{eq:h-main}, i.e., ($h_x + h_y$ ) given explicitly in Eqs. \eqref{hx} and \eqref{hy}, consist of the SOC fields and the applied magnetization (which we have set along the $x$-direction). The energy separation between the Fermi contours of the spin-split bands ($\pm 1$) increases with the magnitude of the spin-dependent terms of the Hamiltonian, i.e., $|h_x\hat{x}+h_y\hat{y}|$. The larger the energy split, the larger would be the $k$-space separation between the contours of the $\pm 1$ Fermi surfaces on which we perform our integration to obtain the spin accumulation $\delta S_y$ (see Fig. 2h), the larger k-space separation translates into a larger net spin accumulation $\delta S_y$. Additionally, $\delta S_y$ is dependent on $h_y$ which in turn arises solely from the SOC terms (noting that $M_y$ is zero), whereas $\delta S_x$ is dependent on $h_x$, which arises from both SOC and the magnetization coupling $M_x$. Hence, the increase in the SOC parameters would lead to a more pronounced increase in $\delta S_y$. The interplay of multiple types of SOC has two important consequences to the net $\delta S_y$: (i) If only a single type of SOC is present, the resulting Sy spin accumulation would be antisymmetric about the Fermi surfaces (refer to Figs. \ref{gFig2}a to \ref{gFig2}c), and would thus cancel out upon integration. This complete cancellation would still occur even in the presence of $M_x$. (ii) When we vary one SOC parameter while keeping the other two fixed, the latter (fixed) SOC terms have their own contributions to $\delta Sy$ and thus provide a constant offset. This is evident from the vertical shifts along the $\delta Sy$ axis of the plots shown in Figs. \ref{gFig4}c and \ref{gFig4}d.

 For the sign conventions adopted for $H_\alpha$, $H_{\beta_3}$, and $H_{\eta3}$ here, $\delta S_y$ becomes more positive when $\eta_3$ and $\alpha$ become more positive and $\eta_3$ becomes more negative. As expected from the similarity of the $\Phi_b(\phi)$ factors between the RTA and SL approaches shown in Fig. \ref{gFig2}i, the $\delta S_y$ obtained using the two approaches are also similar, as shown in the comparison between the two approaches in Fig. \ref{gFig4}c and d, and in Fig. S2c and S2d. This indicates that the RTA is an adequate approximation for investigating the spin accumulation. We now turn our attention to the spin current.

\subsection{Spin current}  

\begin{figure}[htp]
\centering
\includegraphics[width=0.7\textwidth]{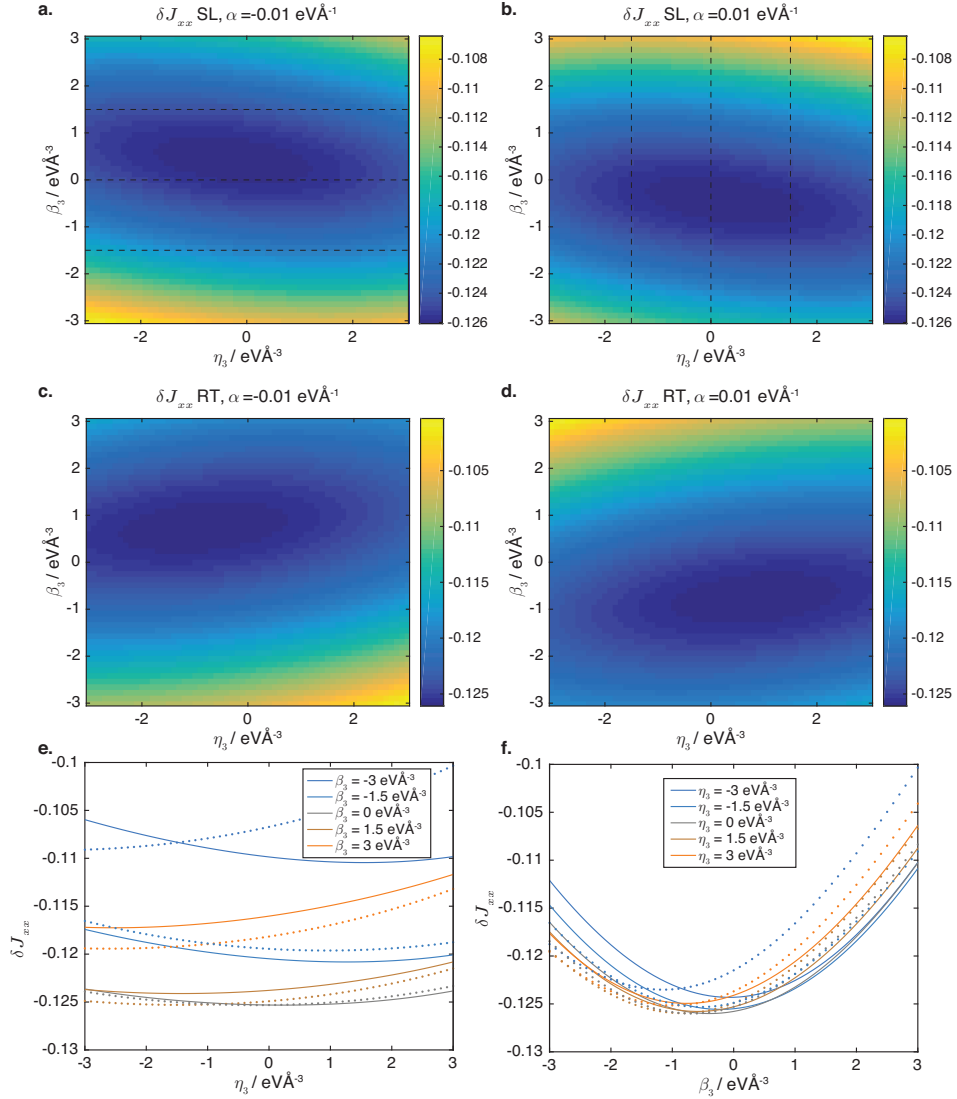}
\caption{\textbf{a.}--\textbf{c.}. $\delta J_{xx}$ as a function of $\beta_3$ and $\eta_3$ obtained using the a.,b. SL and c., d., RTA approaches at a., c. $\alpha=-10$ meV\AA$^{-1}$ and b., d. $\alpha=10$ meV\AA$^{-1}$. \textbf{e}. $\delta J_{xx}$ as a function of $\eta_3$ for the parameters in a. and c. at the values of $\beta_3$ denoted by the dotted lines in a., and \textbf{f}. $\delta J_{xx}$ as a function of $\beta_3$ for the parameters in b. and d. at the values of $\eta_3$ denoted by the dotted lines in b.}
\label{gFig6}
\end{figure}	

Fig. \ref{gFig6} shows $\delta J_{xx}$, i.e., the spin $x$ current flowing along the $x$ direction $\delta J_{xx}$ as functions of $\beta_3$ and $\eta_3$ at two values of $\alpha$ obtained using the SL and RTA approaches. The SL and RTA values of $\delta J_{xx}$ at the specific values of $\beta_3$ which ar denoted by the horizontal dotted lines in Fig. \ref{gFig6}a and specific values of $\eta_3$ indicated by the vertical dotted lines in Fig. \ref{gFig6}b, are plotted in Figs. \ref{gFig6}e and \ref{gFig6}f, respectively, to depict their variational trends. (Note that $J_{xy}$, i.e., the spin $y$ current flowing in the $x$ direction is zero 0 when both the magnetization and electric field are along the $x$ direction.)  Unlike $S_x$, which varies in an approximately linear manner with $\alpha$, $\beta_3$, and $\eta_3$, $\delta J_{xx}$ has an approximately quadratic dependence on $\beta_3$ and $\eta_3$, as shown in Fig. \ref{gFig6},  which results in a non-monotonic variation of $\delta J_{xx}$ with both parameters. (The quadratic dependence is most easily seen in Fig. \ref{gFig6}f where the $\delta J_{xx}$ minima occur within the range of $\beta_3$ considered. The quadratic dependence of $\delta J_{xx}$ on $\eta_3$ is also evident from the minima corresponding to the $\beta_3=0$ eV\AA$^{-3}$ and $\beta_3=0.2$ eV\AA$^{-3}$ lines in Fig. \ref{gFig6}e. )   This quadratic dependence may be due to the fact that $\delta J_{xx}$ involves both the velocity as well as the spin of the carriers, the respective operators of which are combined together in an anti-commutator [Eq. \eqref{eq:OJs}]. Extrema in $\delta J_{xx}$ would then occur because of the competition between the opposing trends experienced by the two quantities (velocity and spin accumulation) Comparing Figs. \ref{gFig6}a and \ref{gFig6}b, it is evident that the maximum amplitude of the spin current occurs at relatively small values of $\beta_3$ and $\eta_3$.  Furthermore, the signs of $\beta_3$ and $\eta_3$ at which this maximum amplitude occurs (positive $\beta_3$ and negative $\eta_3$ at $\alpha=-0.01$ eV\AA$^{-1}$ in Fig. \ref{gFig6}a and vice-versa at $\alpha=0.01$ eV\AA$^{-1}$ in Fig. \ref{gFig6}b) can change with the value of $\alpha$. A comparison between Figs. \ref{gFig6}a and \ref{gFig6}c, and between Figs. \ref{gFig6}b and \ref{gFig6}d shows that unlike the spin $y$ accumulation $\delta S_y$, for which both the SL and RTA approaches give similar results, there are now evident differences between the $\delta J_{xx}$ profiles obtained using the SL approach in Figs. \ref{gFig6}a and \ref{gFig6}b, as opposed to the RTA approach in Figs. \ref{gFig6}c and \ref{gFig6}d. These differences are due to the displacement of the SL and RTA quadratic $\delta J_{xx}$ vs $\beta_3$ (Fig. \ref{gFig6}e) and $\delta J_{xx}$ vs $\eta_3$ curves (Fig. \ref{gFig6}f) curves with respect to each other, which can be in turn attributed to the anisotropy of the SLA relaxation time on the Fermi contours (see discussion on Fig. S3 in Supplementary Materials). The discrepancy between the SL and RTA results for the spin current indicates that the RTA may not be an adequate approximation for the spin current although the main qualitative trends are still captured correctly. We therefore show only the SL values for the subsequent results. 

\begin{figure}[htp]
\centering
\includegraphics[width=0.7\textwidth]{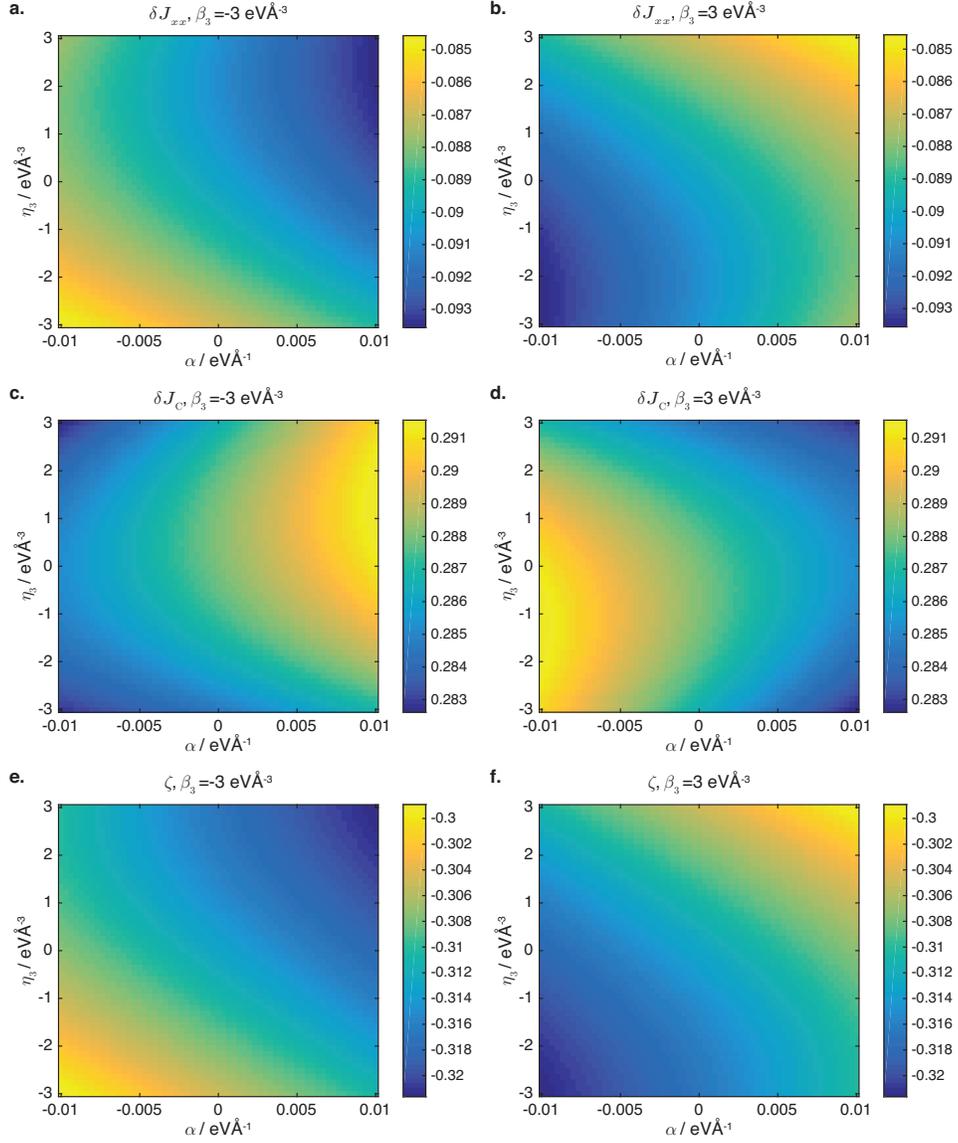}
\caption{\textbf{a.},\textbf{b.} $\delta J_{xx}$ as a function of $\alpha$ and $\beta_3$ obtained at a. $\beta_3=-3$ meV\AA$^{-3}$ and b. $\beta_3=3$ meV\AA$^{-3}$. \textbf{c},\textbf{d}. The corresponding charge current $\delta J_{\mathrm{C}}$ flowing along the $x$ direction, and \textbf{e}, \textbf{f} charge-spin efficiency $\xi\equiv \delta J_{xx}/\delta J_{\mathrm{C}}$.}

\label{gFig7}
\end{figure}	

The charge-spin conversion efficiency \cite{NatPhy11_496, NatMat15_1261, PRL119_077702} , i.e., the ratio of the spin current to the charge current, is an important figure of merit in technological applications. Fig. \ref{gFig7} shows $\delta J_{xx}$, the corresponding charge current $\delta J_{\mathrm{C}}$, and the spin charge efficiency $\xi\equiv \delta J_{xx}/\delta J_{\mathrm{C}}$ plotted as functions of $\alpha$ and $\eta_3$ at two values of $\beta_3$. $\delta J_{xx}$ is plotted as a function of $\alpha$ and $\eta_3$ here to complement to the plots of $\delta J_{xx}$ as a function of $\beta_3$ and $\eta_3$ to provide snapshots of the variation of $\delta J_{xx}$ in Fig. \ref{gFig6}, while $\delta J_{xx}$, $\delta J_{\mathrm{C}}$, and $\xi$ are plotted for two values of $\beta_3$ here to show that the qualitative behavior of $\delta J_{xx}$ are similar across different values of $\beta_3$. (The corresponding plots of $\delta J_{\mathrm{C}}$ and $\xi$ for the parameter sets of Fig. \ref{gFig6} are shown in Fig. S4 in the Supplementary Materials.) Similar to its dependence on $\beta_3$ and $\eta_3$, $\delta J_{xx}$ also has a quadratic dependence on $\alpha$ although the quadratic minima do not fall within the ranges of $\alpha$ considered here. Comparing  the $\delta J_{xx}$ values in Fig. \ref{gFig7}a and b against the $J_{\mathrm{C}}$ values in Fig. \ref{gFig7}c and d, it is evident that the regions on the $\alpha-\eta_3$ plane at which $\delta J_{xx}$ has large magnitudes at large positive (negative) values of $\eta_3$ and $\alpha$ at $\beta_3=-3$ eV\AA$^{-3}$ ($\beta_3=3$ eV\AA$^{-3}$) approximately coincide with the regions of small magnitudes of $\delta J_{\mathrm{C}}$ on the $\alpha-\eta_3$ plane. Fig. S4 in the Supplementary Materials shows that this coincidence between the regions of large $\delta J_{xx}$ and small $\delta J_{\mathrm{C}}$ holds on the $\beta_3-\eta_3$ plane as well. The large values of $J_{\mathrm{xx}}$ with small values of charge current translate into a large charge-spin conversion efficiency in excess of 30\%.

\section{Conclusion}
In this work, we investigated how the interplay between the three types of RSOC in a LaO/STO system affects the spin accumulation and spin current when a magnetization and an electric field are applied. The spin accumulation and spin currents were calculated using both the RTA and the Schliemann-Loss approach. 

For the spin accumulation, we found that within the parameter ranges investigated, the spin accumulation in the direction perpendicular to the magnetization varies approximately with each of the three RSOC strengths when the other two RSOC strengths are fixed. The latter two RSOC strengths provide an offset to the value of the varied RSOC at which the spin accumulation changes sign. The spin accumulation, in general, increases when the magnitudes of the three RSOC strengths are increased. The RTA results for the spin accumulation agree adequately with those of the more accurate Schliemann-Loss approach. 

For the spin current, we found that the spin current varies quadratically with each of the three RSOC strengths when the other two RSOC strengths are fixed. The minima of this quadratic dependence depends on the values of the other two RSOC strengths. In contrast to the spin accumulation, there is some variance between the spin current profiles between the results of the RTA and Schliemann-Loss approaches because the anisotropy of the Fermi contours were not accounted for in the former. This suggests that the Schliemannn-Loss approach should be used to calculate the spin currents accurately. A large spin-charge conversion efficiency in excess of 30 \% can be achieved in the system. 

Our results provide a more complete understanding of how the spin response of the low-energy states in the LaO/STO system is affected by the co-existence of multiple types of RSOC. 
     
\section{Acknowledgments}

This work is supported by the Ministry of Education (MOE) Tier-II grants MOE2018-T2-2-117 (NUS Grant Nos. A-0005163-00-00) and MOE-T2EP50121-0014 (NUS Grant Nos. A-8000086-01-00), and MOE Tier-I FRC grants (NUS Grant Nos. A-0005110-01-00 and A-8000195-01-00).

 \bibliographystyle{unsrt}
\bibliography{Bibtex-LAO-STO-v2}

\end{document}


\title{Semiclassical spin-transport in LaO/STO system}

\author{Anirban Kundu}
\email{??}
\selectlanguage{english}%
\affiliation{Department of Electrical and Computer Engineering, National University of Singapore, Singapore}

\author{Zhuo Bin Siu}
\email{elesiuz@nus.edu.sg}
\selectlanguage{english}%
\affiliation{Department of Electrical and Computer Engineering, National University of Singapore, Singapore}

\author{Mansoor B.A. Jalil}
\email{elembaj@nus.edu.sg}
\selectlanguage{english}%
\affiliation{Department of Electrical and Computer Engineering, National University of Singapore, Singapore}
\maketitle
\begin{figure}[htp]
\centering
\includegraphics[width=0.7\textwidth]{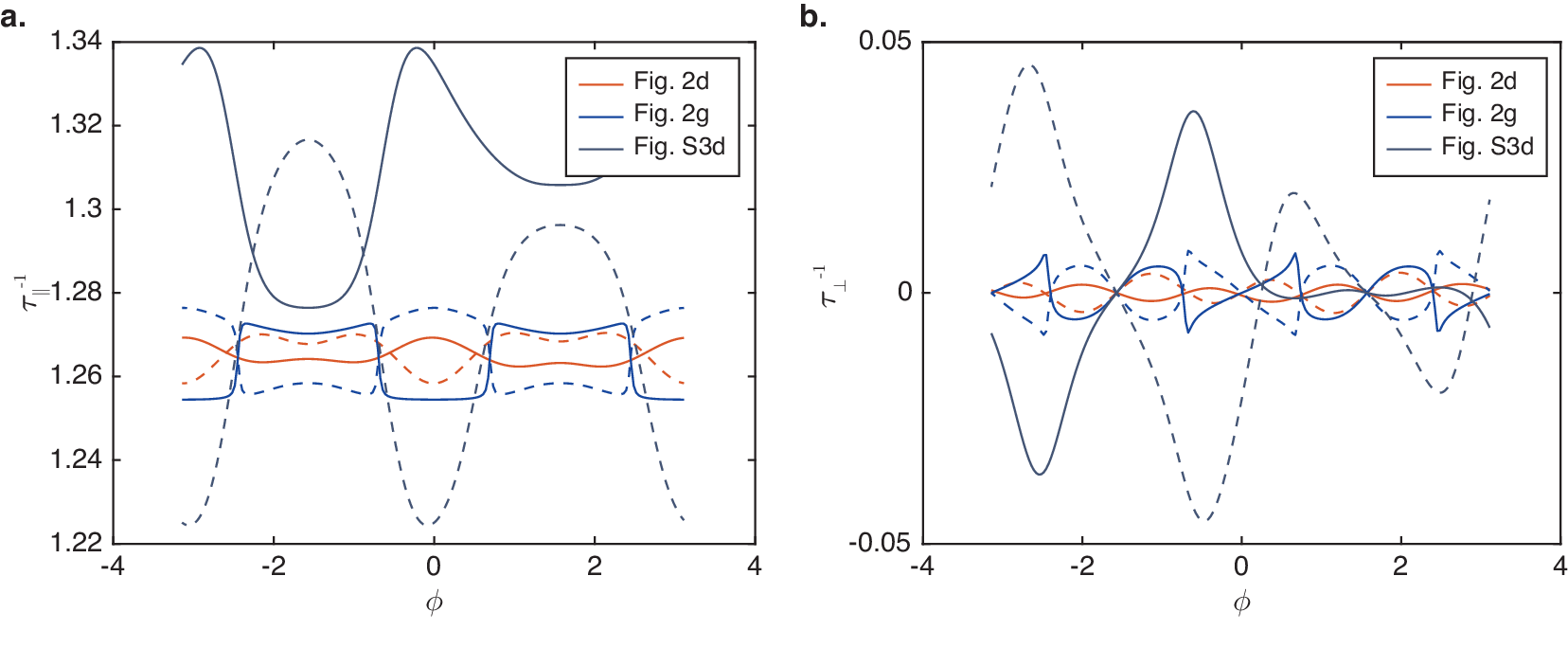}
\caption{ 
Exemplary values of \textbf{a} $1/\tau^\parallel_{\mathbf{k}b}$ and \textbf{b} $1/\tau^\perp_{\mathbf{k}b)}$ for the parameter sets of Fig. 2d ($J_HM=0$ eV, $\alpha=5$ meV\AA$^{-1}$,and $\beta_3=\eta_3=1.5$ eV\AA$^{-3}$) and Fig. 2g ($J_HM=10$ meV, $\alpha=5$ meV\AA$^{-1}$,and $\beta_3=\eta_3=1.5$ eV\AA$^{-3}$) in the main text, and Fig. S3d in the Supplementary Materials ($J_HM=10$ meV, $\alpha=10$ meV\AA$^{-1}$,and $\beta_3=-\eta_3=3$ eV\AA$^{-3}$). The solid and dotted lines correspond to values for the $b=+$ and $b=-$ bands respectively. }
\label{gSuppFig1}
\end{figure}

Fig. \ref{gSuppFig1} shows that $1/\tau^\parallel_{\mathbf{k}b}$ is typically much larger than $1/\tau^\perp_{\mathbf{k}b}$ for a few parameter sets used in the figures on this paper. (Note the difference in the $y$ axis scales of the two panels. ) The contribution of $1/\tau^\perp_{\mathbf{k}b}$ in the denominator of Eq. (31), and the contribution of the $f^{(1)\perp}_{\mathbf{k}b}$ in Eq. (32) may therefore be neglected to a first approximation.

\begin{figure}[htp]
\centering
\includegraphics[width=0.7\textwidth]{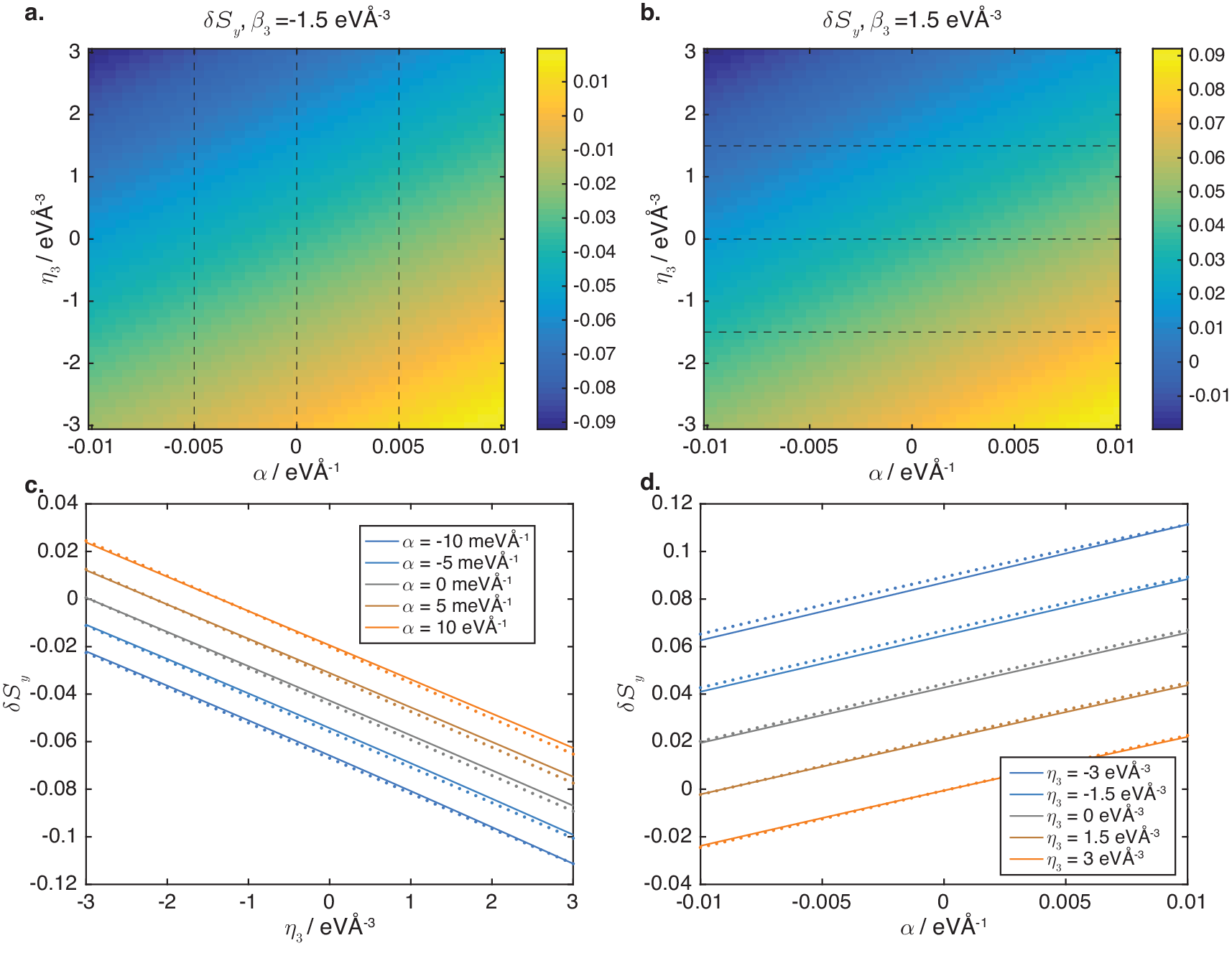}
\caption{ 
\textbf{a},\textbf{b} Spin $y$ accumulation $\delta S_y$ as a function of $\eta_3$ and $\alpha_3$ at a. $\beta_3=-1.5$ eV\AA$^{-3}$ and b. $\beta_3=1.5$ eV\AA$^{-3}$. \textbf{c} $\delta S_y$ as a function of $\eta_3$ for the parameters in a. at the various values of $\alpha$ indicated by the the dotted lines in a., and \textbf{d} $\delta S_y$ as a function of $\alpha$ for the parameters in b. a the various values of $\eta_3$ indicated by the dotted lines in b. The solid lines in c. and d. denote $\delta S_y$ obtained using the SL approach and the dots $\delta S_y$ obtained using the RTA. }
\label{gFig5}
\end{figure}	

Fig. \ref{gFig5} is the counterpart of Fig. 4 in the main text for which the spin $y$ accumulation is now plotted as functions of $\eta_3$ and $\alpha_3$ here to show that $\delta S_y$ has an approximately linear dependence on $\eta_3$ and $\alpha$ and that the values of $\delta S_y$ obtained using the SL and RTA approaches are similar to each other.

\begin{figure}[htp]
\centering
\includegraphics[width=0.7\textwidth]{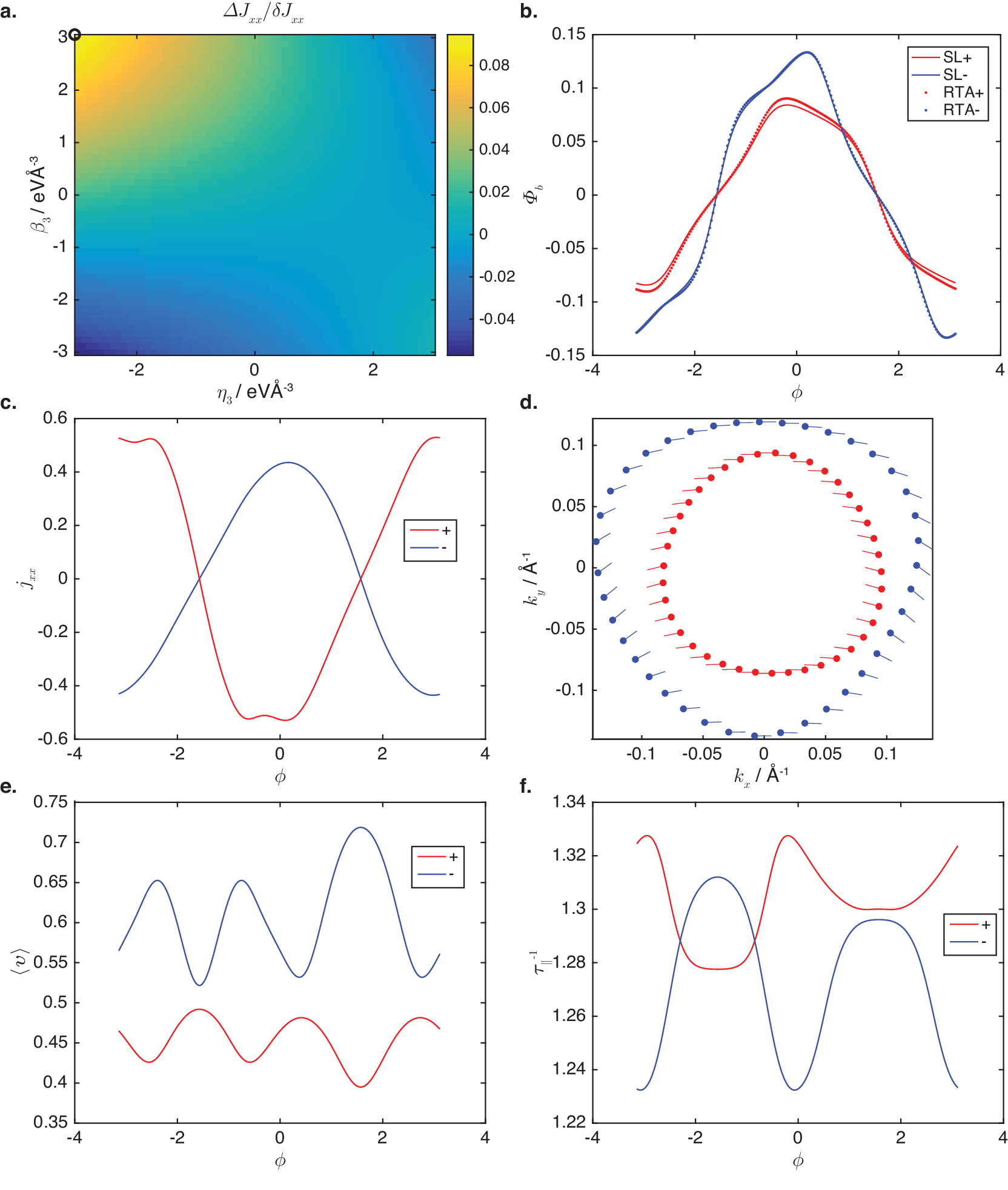}
\caption{  
\textbf{a.} The relative discrepancies between the SL and RTA values for $\delta J_{xx}$ as a function of $\beta_3$ and $\eta_3$ at $\alpha=10$ meV\AA$^{-1}$. \textbf{b}. $\mathrm{PM}$ obtained using SL and RTA for the two bands at $\beta_3=-\eta_3=3$ eV\AA$^{-1}$ indicated by the circle in a. \textbf{c}. The expectation current of the spin $x$ current flowing along $x$ $\delta J_{xx}$ for each of the states on the two Fermi contours as a function of $\phi$. \textbf{d}. The Fermi contours of the two bands and the spin expectation value vectors for each state on the Fermi contours. \textbf{e} Expectation value of velocity on the two Fermi contours as functions of $\phi$  and \textbf{f} the inverse of $\tau^{\parallel}_{\mathbf{k}b}$. }
\label{gFig10}
\end{figure}	

Fig. \ref{gFig10} provides a closer look at the discrepancy between the values of $\delta S_y$ obtained using the SL and RTA approaches in Fig. 5 of the main text. Fig. \ref{gFig10}a shows the relative discrepancy between the SL and RTA values of $\delta S_y$ for the parameter set of Fig. 5b  in which the discrepancy is defined as $2(\delta J_{xx}^{\mathrm{SL}} - \delta J_{xx}^{\mathrm{RTA}})/(\delta J_{xx}^{\mathrm{SL}} + \delta J_{xx}^{\mathrm{RTA}})$. Within the parameter range considered, the discrepancy has the largest magnitude at $\beta_3=-\eta_3 = 3$ eV\AA$^{-3}$, which is circled in Fig. \ref{gFig10}a.  We therefore zoom at this set of $\beta_3$ and $\eta_3$. The immediate reason for the discrepancy between the SL and RTA approaches at this set of $\beta_3$ and $\eta_3$ can be traced to the discrepancies between the $\Phi_b(\phi)$ prefactors for the two approaches, which are plotted in Fig. \ref{gFig10}b. Unlike Fig. 2f and 2g in the main text, the differences between the SL and RTA $\Phi_b(\phi)$ values respectively represented by the dotted lines and dots are now visible on the scale of the figure. The difference between the SL and RTA values of $\delta J_{xx}$ is further exacerbated by the fact that the values of $\phi$ at which there are large differences between the SL and RTA $\Phi_b(\phi)$s around $\phi=0$ and $\phi=\pi$ coincide with the values of $\phi$ at which $\delta J_{xx}(\phi)$ has large magnitudes (Fig. \ref{gFig10}c). The differences between the SL and RTA values of $\Phi_b(\phi)$ can in turn be traced to the anisotropy of the Fermi contours, which are shown in Fig. \ref{gFig10}d. Compared to the Fermi contours shown for a smaller magnitudes of all three SOI strengths in Fig. 2g in the main text, the Fermi contours in Fig. \ref{gFig10}e show a larger deviation from the symmetric elliptical shapes and their centers are further displaced from each other. The anisotropy of the Fermi contours leads to a large $\phi$ angular dependence of the various terms involved in the calculation of $\tau^\parallel$ in Eq. 33 in the main text, including the velocities, which are plotted in Fig. \ref{gFig10}e. This results in a variation of $\tau^\parallel_{\mathbf{k}b}$ by up to almost 5\% with the variation of $\phi$ (Fig. \ref{gFig10}f) , which is a considerable deviation from the assumption that $\tau$ is independent of $\phi$in the RTA approach. In comparison, the variation of $\tau^\parallel_{\mathbf{k}b}$ for the other parameter sets shown in Fig. \ref{gSuppFig1}, for which the Fermi contours are less anisotropic,  are relatively smaller. Interestingly, the value of $\eta_3$ and $\beta_3$ corresponds to that at which the largest magnitude of spin $y$ accumulation occurs (Fig. 5b in the main text). This is not unexpected because, as explained at the end of Sect. 3.1 in the main text, a larger anisotropy in the Fermi contour is correlated with a spin $y$ expectation value of the states on the Fermi contour owing to the larger influence of the stronger SOC effect relative to the fixed magnetization exchange coupling strength.  A second contributory factor to the discrepancy between the SL and RTA results is that the contributions of terms proportional to $Ev_y$ in $f^{(1)\perp}_{\mathbf{k}b}$ (Eq. (32) in the main text) are neglected in the RTA approach. This results in larger discrepancies when $(1/\tau^{\perp}_{\mathbf{k}b})$ has larger values, as is the case here when $\beta_3=-\eta_3 = 3$ eV\AA$^{-3}$ (Fig. \ref{gSuppFig1}b).   A further contributory factor is that the same value of $\tau$ is assumed for both bands in the RTA approach, whereas Fig. \ref{gFig10}f shows that the values of $\tau^\parallel$ can differ by up to almost 8\% between the two bands. 

\begin{figure}[htp]
\centering
\includegraphics[width=0.7\textwidth]{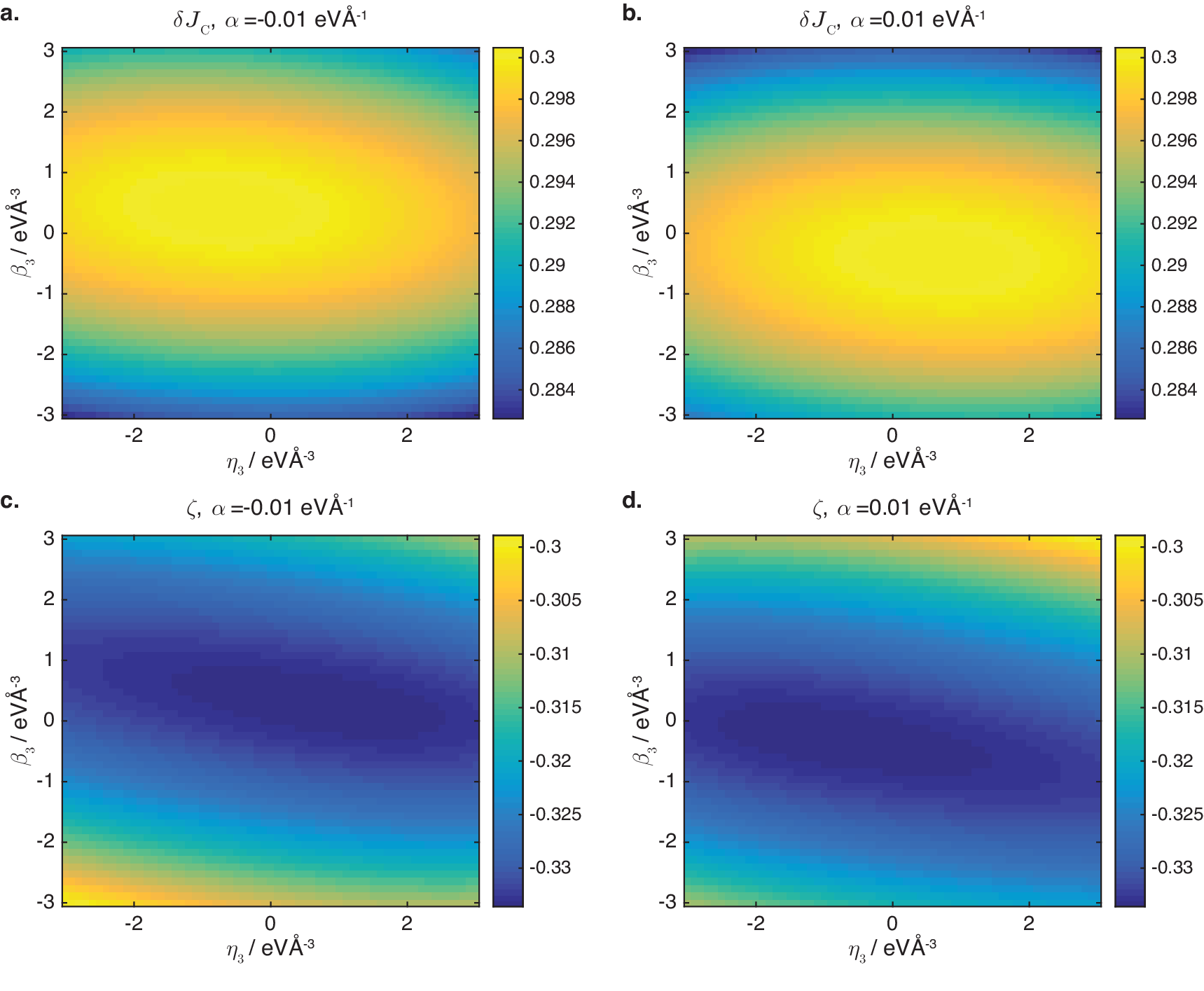}
\caption{  
\textbf{a.},\textbf{b.}$\delta J_{\mathrm{C}}$ as a function of $\beta_3$ and $\eta_3$ at a. $\alpha=-10$ meV\AA$^{-1}$ and b. $\alpha=10$ meV\AA$^{-1}$, \textbf{c.}, \textbf{d.} the corresponding charge-spin efficiency $\eta\equiv \delta J_{xx}/\delta J_{\mathrm{C}}$.}
\label{gFig8}
\end{figure}	

Fig. \ref{gFig8} shows the charge current and charge-spin conversion efficiency for the parameter settings of Fig. 5 of the main text as the $\beta_3-\eta_3$ counterpart of the $\eta_3-\alpha$ plane results in Fig. 6 of the main text. A similar quadratic variation of the charge current with $\beta_3$ and $\eta_3$ and coincidence between the regions of large $\delta J_{xx}$ and small $\delta J_{\mathrm{C}}$ to those in Fig. 6 of the main text are observed.